\documentclass[twocolumn]{aastex631}
\usepackage{graphicx}
\usepackage{comment}
\usepackage{amsmath, amssymb}
\usepackage{bm}
\usepackage{hyperref}
\usepackage{soul}
\usepackage{upgreek}
\usepackage{enumitem}
\usepackage{natbib}
\usepackage{tabularx}
\usepackage{float}
\usepackage[normalem]{ulem}
\usepackage{gensymb}

\begin{document}

\title{A Heavily Scattered Fast Radio Burst Is Viewed Through Multiple Galaxy Halos}

\author[0000-0001-9855-5781]{Jakob T. Faber}
\affiliation{Cahill Center for Astronomy and Astrophysics, MC 249-17 California Institute of Technology, Pasadena CA 91125, USA.}

\author[0000-0002-7252-5485]{Vikram Ravi}
\affiliation{Cahill Center for Astronomy and Astrophysics, MC 249-17 California Institute of Technology, Pasadena CA 91125, USA.}
\affiliation{Owens Valley Radio Observatory, California Institute of Technology, Big Pine CA 93513, USA.}

\author[0000-0002-4941-5333]{Stella Koch Ocker}
\affiliation{Cahill Center for Astronomy and Astrophysics, MC 249-17 California Institute of Technology, Pasadena CA 91125, USA.}
\affiliation{The Observatories of the Carnegie Institution for Science, Pasadena, CA 91101, USA.}

\author[0000-0002-6573-7316]{Myles B. Sherman}
\affiliation{Cahill Center for Astronomy and Astrophysics, MC 249-17 California Institute of Technology, Pasadena CA 91125, USA.}

\author[0000-0002-4477-3625]{Kritti Sharma}
\affiliation{Cahill Center for Astronomy and Astrophysics, MC 249-17 California Institute of Technology, Pasadena CA 91125, USA.}

\author[0000-0002-7587-6352]{Liam Connor}
\affiliation{Cahill Center for Astronomy and Astrophysics, MC 249-17 California Institute of Technology, Pasadena CA 91125, USA.}

\author[0000-0002-4119-9963]{Casey Law}
\affiliation{Cahill Center for Astronomy and Astrophysics, MC 249-17 California Institute of Technology, Pasadena CA 91125, USA.}
\affiliation{Owens Valley Radio Observatory, California Institute of Technology, Big Pine CA 93513, USA.}

\author[0000-0003-1226-118X]{Nikita Kosogorov}
\affiliation{Cahill Center for Astronomy and Astrophysics, MC 249-17 California Institute of Technology, Pasadena CA 91125, USA.}

\author[0000-0002-7083-4049]{Gregg Hallinan}
\affiliation{Cahill Center for Astronomy and Astrophysics, MC 249-17 California Institute of Technology, Pasadena CA 91125, USA.}
\affiliation{Owens Valley Radio Observatory, California Institute of Technology, Big Pine CA 93513, USA.}

\author{Charlie Harnach}
\affiliation{Owens Valley Radio Observatory, California Institute of Technology, Big Pine CA 93513, USA.}

\author[0000-0002-8191-3885]{Greg Hellbourg}
\affiliation{Cahill Center for Astronomy and Astrophysics, MC 249-17 California Institute of Technology, Pasadena CA 91125, USA.}
\affiliation{Owens Valley Radio Observatory, California Institute of Technology, Big Pine CA 93513, USA.}

\author{Rick Hobbs}
\affiliation{Owens Valley Radio Observatory, California Institute of Technology, Big Pine CA 93513, USA.}

\author{David Hodge}
\affiliation{Cahill Center for Astronomy and Astrophysics, MC 249-17 California Institute of Technology, Pasadena CA 91125, USA.}

\author{Mark Hodges}
\affiliation{Owens Valley Radio Observatory, California Institute of Technology, Big Pine CA 93513, USA.}

\author[0000-0002-5959-1285]{James W. Lamb}
\affiliation{Owens Valley Radio Observatory, California Institute of Technology, Big Pine CA 93513, USA.}

\author{Paul Rasmussen}
\affiliation{Owens Valley Radio Observatory, California Institute of Technology, Big Pine CA 93513, USA.}

\author[0000-0001-8426-5732]{Jean J. Somalwar}
\affiliation{Cahill Center for Astronomy and Astrophysics, MC 249-17 California Institute of Technology, Pasadena CA 91125, USA.}

\author[0000-0002-9353-6204]{Sander Weinreb}
\affiliation{Cahill Center for Astronomy and Astrophysics, MC 249-17 California Institute of Technology, Pasadena CA 91125, USA.}

\author{David P. Woody}
\affiliation{Owens Valley Radio Observatory, California Institute of Technology, Big Pine CA 93513, USA.}

\collaboration{200}{(The Deep Synoptic Array team)}

\begin{abstract}
We present a multi-wavelength study of the apparently non-repeating, heavily scattered fast radio burst, FRB\,20221219A, detected by the Deep Synoptic Array 110 (DSA-110). The burst exhibits a moderate dispersion measure (DM) of $706.7^{+0.6}_{-0.6}$ $\mathrm{pc}~\mathrm{cm}^{-3}$ and an unusually high scattering timescale of $\tau_{\mathrm{obs}} = 19.2_{-2.7}^{+2.7}$ ms at 1.4 GHz. We associate the FRB with a Milky Way-like host galaxy at $z_{\mathrm{host}} = 0.554$ of stellar mass $\mathrm{log}_{10}(M_{\star, \mathrm{host}}) = 10.20^{+0.04}_{-0.03} ~M_\odot$. We identify two intervening galaxy halos at redshifts $z_{\mathrm{igh1}} = 0.492$ and $z_{\mathrm{igh2}} = 0.438$, with low impact parameters, $b_{\mathrm{igh1}} = 43.0_{-11.3}^{+11.3}$ kpc and $b_{\mathrm{igh2}} = 36.1_{-11.3}^{+11.3}$ kpc, and intermediate stellar masses, $\mathrm{log}_{10}(M_{\star, \mathrm{igh1}}) = 10.01^{+0.02}_{-0.02} ~M_\odot$ and $\mathrm{log}_{10}(M_{\star, \mathrm{igh2}}) = 10.60^{+0.02}_{-0.02} ~M_\odot$. The presence of two such galaxies suggests that the sightline is significantly overcrowded compared to the median sightline to this redshift, as inferred from the halo mass function. We perform a detailed analysis of the sightline toward FRB\,20221219A, constructing both DM and scattering budgets. Our results suggest that, unlike most well-localized sources, the host galaxy does not dominate the observed scattering. Instead, we posit that an intersection with a single partially ionized cloudlet in the circumgalactic medium of an intervening galaxy could account for the substantial scattering in FRB\,20221219A and remain in agreement with typical electron densities inferred for extra-planar dense cloud-like structures in the Galactic and extragalactic halos (e.g., high-velocity clouds).
\end{abstract}


\section{Introduction}

Fast radio bursts \citep[FRBs;][]{Lorimer2007} are a class of highly luminous extragalactic radio transients with durations ranging from nanoseconds to seconds \citep{Petroff2019, Cordes2019, Majid2021, Nimmo2022, chime2022}. Their dispersion measures (DMs), which quantify the electron column density along the line of sight (LoS), exceed the values expected from Milky Way electron density models, suggesting extragalactic origins. This hypothesis has since been confirmed by an increasing number of well-localized FRBs that reside in host galaxies out to redshifts z $\sim$ 1 \citep{Ryder2023, Law2023, Gordon2023}.

While the physical origins of FRBs remain unknown, many source models have been proposed, ranging from isolated neutron stars to compact object mergers. Recent observations have noted magnetars as viable sources, supported by detections of millisecond radio bursts from the galactic magnetar SGR\,1935$+$2154 by the STARE-2 and CHIME/FRB observatories, with measured isotropic-equivalent energies of $2.2^{+0.4}_{-0.4} \times 10^{35} ~\mathrm{erg}$ and $3_{-1.6}^{+3} \times 10^{35} ~\mathrm{erg}$, respectively, consistent with FRB source energetics \citep{Bochenek2020, chime2020b}. 

Despite their unknown formation channels, sources, and emission mechanisms, if the source distance is known and intervening material is well-characterized, FRB DMs serve as stringent probes of baryon densities within galaxy halos, clusters, and filaments throughout the intergalactic medium \citep[IGM;][]{McQuinn2014, Macquart2020}. FRB DMs can also facilitate the study of diffuse ionized gas in the Milky Way and neighboring galaxies \citep{Connor2022,Ravi2023a, Cook2023}.

FRB observations are sensitive to inhomogeneities in intervening plasma through the effects of multipath propagation, or scattering. This propagation effect results in asymmetric, frequency-dependent pulse broadening and broadly exponential scattering tails (assuming a Gaussian scattered image). In the case of interstellar scattering, such tails are expected to follow a temporal delay $\tau \propto \nu^{-\alpha}$, where $\alpha \gtrsim 4$ assuming an infinitely wide, thin scattering screen described by Kolmogorov turbulence \citep[$\alpha = 4.4$ for scale-free power spectra, $\alpha = 4$ in the case of a diffractive scale below a finite inner scale, or where the density fluctuations follow a Gaussian power spectrum;][]{Rickett2009}. These assumptions are largely idealized, however, and may be challenged by flatter values ($\alpha \lesssim 4$) inferred for Galactic pulsars \citep{Lohmer2001, Deneva2009, Dexter2017}. The diffractive scintillation bandwidth $\Delta \nu_{\mathrm{d}} \simeq(2 \pi \tau)^{-1}$ yields more information due to its preservation of phase information, though in practice, most of the convincing measurements have been made for scintillation caused by Galactic scattering \citep{Masui2015, Gajjar2018, Hessels2019, Marcote2020, Bhandari2020, Schoen2021, Ocker2022a, Sammons2023}, with the exception of FRB\,150807 \citep{Ravi2016}. While some FRBs at lower Galactic latitudes (e.g., FRB\,20121102A, FRB\,20180916B) are predominantly scattered by the Milky Way (MW) \citep{Ocker2021}, the majority exhibit scattering timescales ($\tau(\nu)$) that exceed those expected from purely Galactic contributions \citep{Cordes2002}. The dominant scattering source for these FRBs is unconstrained in the absence of both scintillation bandwidth and scattering timescales, which can be used in tandem to constrain the presence of both Galactic and extragalactic screens \citep{Cordes2016}.

To date, sightline-independent correlations between DM and $\tau$ have been explored for the broader FRB population \citep{Ravi2018, chime2019}, and compared to the known $\tau$-DM correlation observed for MW radio pulsars \citep{Cordes2016, Cordes2022}. Such comparisons have been difficult, however, as the canonical $\tau$-DM relation for pulsars relies on Galactic DMs, while the DMs typically assumed for FRBs in comparison are extragalactic (DM$_{\mathrm{ex}}$; beyond the MW) and not specific to the host galaxy of the source. The relatively small sample of well-localized FRBs available has limited our ability to precisely constrain host DMs, and make what would be more apt comparisons between $\tau$ and DM$_{\mathrm{host}}$, rather than DM$_{\mathrm{ex}}$. However, the recent advent of precise FRB localizations has begun alleviating this dilemma, which we discuss in \S\protect\ref{sec:dmbudget}. 

Recent studies favor either the host galaxy or circumburst environment as the dominant scattering medium \citep{Simha2020, Chittidi2021, Ocker2022b, Cordes2022}, though it has been suggested that intervening halos may contribute meaningfully to scattering in select cases as well \citep{Vedantham2018, Connor2020, Chawla2022}. Determining whether scattering is, in certain instances, dominated by the circumburst medium, host, or circumgalactic media (CGM) of intervening galaxies could further constrain source models and offer insights into these galactic halos, opening avenues for detailed studies of plasma turbulence in both the host and intervening galaxies \citep{Cordes2016, Simard2021, Ocker2022b}.

For scattering to dominate in the circumburst medium, high electron densities $n_{e} \gg 10^{-2}$ cm$^{-3}$ and high-amplitude density fluctuations \citep[far exceeding a MW-like ISM;][]{Ocker2020} would be expected, as well as strong magnetic fields $B \gtrsim 50 ~\mu \mathrm{G}$ \citep[similar to the magnetized filaments observed in the Crab nebula, for instance;][]{Bietenholz1991} that induce extreme Faraday rotation. Highly magnetic environments do not, however, guarantee strong scattering \citep[e.g., in the case of the first repeater FRB\,20121102A;][]{Michilli2018}. To date, evidence for circumburst scattering has only been found in FRB\,20190520B, which exhibits rapid variability in scattering between bursts \citep{Ocker2022b}.

Significant scattering in the interstellar medium (ISM) of FRB host galaxies appears to be more common \citep{Masui2015,Ocker2022a}. Such constraints require contemporaneous and discrepant scintillation and scattering measurements, such that a two-screen scattering model can be applied \citep{Simard2021}. The two-screen model does, however, come with certain assumptions: e.g., that there are only two dominant scattering screens. Furthermore, placing constraints on the model necessitates measurable scintillation in the scattering tail of the burst. Scattering in one or more intervening galaxies, for instance, further complicates the two-screen model. 


In this paper, we analyze the contributions to both DM and scattering in FRB\,20221219A by intervening plasmas. In \S\protect\ref{sec:observations} we present the detection of the heavily scattered burst, including its localization with the DSA-110 and optical/IR follow-up observations of its host galaxy and two intervening galaxies. In \S\protect\ref{sec:dmbudget} we constrain contributions to the DM along the LoS (the so-called ``DM budget'') using HI-inferred HII column densities for intervening galaxies based on results from the COS-Halos Survey \citep{Werk2014}, and Illustris TNG simulation data \citep{Zhang2021} for the intergalactic medium (IGM). In \S\protect\ref{sec:scatbudget} we present a detailed investigation aimed at identifying the medium responsible for the uniquely high degree of scattering (the so-called ``scattering budget''), with a focus on the circumgalactic media (CGMs) of two intervening galaxies. We conclude in \S\protect\ref{sec:futurework}. In this work, we adopt standard cosmological parameters from \citet{Planck2015} for estimating DM contributions from the IGM \citep[consistent with Illustris TNG;][]{Pillepich2017} and from \citet{Planck2020} for all other cosmological inferences.

\section{DSA-110 Observation of FRB\,20221219A} \label{sec:observations}

\begin{table}
    \caption{Burst properties for FRB\,20221219A (see Figure~\ref{fig:nihariwfall}).}
    \label{tab:nihariprops}
    \centering
    \hspace*{-0.97cm}
    \resizebox{0.53\textwidth}{!}{
        \begin{tabularx}{0.55\textwidth}{@{\extracolsep{\fill}}XX}
            \hline \hline 
            Parameter & Value \\
            \hline 
            $\mathrm{Localization ~[R.A. ~J2000, ~Dec. ~J2000]}$ & $17^{h}10^{m}31.15^{s}, +71^{\circ}37\arcmin36.6\arcsec$ \\
            $\mathrm{Localization ~Uncertainty ~[1\sigma, arcsec]}$ & 1.5, 0.9 \\
            $\mathrm{Signal}\text{-}\mathrm{to}\text{-}\mathrm{Noise}$ & $8.9^{+1}_{-1}$ \\
            $\mathrm{Dispersion ~Measure ~[pc ~cm^{-3}]}$ & $706.7^{+0.6}_{-0.6}$ \\
            $\mathrm{Intrinsic ~FWHM ~[ms]}$ & $0.12_{-0.08}^{+0.65}$ \\
            $\mathrm{Scattering ~Timescale ~[ms]}$ & $19.2_{-2.7}^{+2.7}$ \\
            $\mathrm{Linear ~Polarization ~Fraction ~[L/I]}$ & $0.502_{0.08}$ \\
            \hline
        \end{tabularx}
    }
\end{table}

\subsection{DSA-110 Discovery and Localization}

As described in \citet{Ravi2023b}, the DSA-110 is a radio interferometer located at the Owens Valley Radio Observatory (OVRO) dedicated to the discovery and arcsecond-scale localization of FRBs. FRB\,20221219A was detected during commissioning observations at a Modified Julian Date (MJD) of 59932.79297813 (arrival time at 1530\,MHz at the observatory), with a real-time detection signal-to-noise ratio\footnote{This was calculated using a weighted sum matched filter.} of 8.9. Antenna voltage measurements containing the burst were recorded in two linear polarizations with a time resolution of 32.768 $\mu s$ across 6144 channels between 1311.25--1498.75\,MHz. These data were used to localize the burst to a sky location of (R.A. J2000, Dec. J2000) = ($+17^{h}10^{m}31.15^{s}, +71^{\circ}37\arcmin36.6\arcsec$) with $1\sigma$ uncertainties of 1.5\arcsec~in R.A. and 0.9\arcsec~in Dec. The data were also coherently combined to measure the various burst properties given in Table~\ref{tab:nihariprops}, and a dynamic spectrum of the burst is shown in Figure~\ref{fig:nihariwfall}. 

\subsection{Inference of Burst Properties}

We optimize the DM for temporal structure in the burst by searching for the DM at which the forward-derivative of the dedispersed timeseries is maximized, following methods outlined in \cite{Gajjar2018}, \cite{Hessels2019}, and \cite{Josephy2019}. We begin by calculating a 2D DM transform across the burst in time for 0.075 $\mathrm{pc}~\mathrm{cm}^{-3}$ intervals, ranging from $692~\text{-}~722$ $\mathrm{pc}~\mathrm{cm}^{-3}$. We then calculate the forward-derivative and convolve with a $3 \times 5\left(2.7~\mathrm{ms} \times 0.5 ~\mathrm{pc} ~\mathrm{cm}^{-3}\right)$ smoothing kernel. Finally, we integrate the modulus of the forward-derivative in time raised to $n \in(1,2,4)$. \cite{Gajjar2018} and \cite{Hessels2019} elect to use $n=1$ and $n=2$, respectively. \cite{Josephy2019}, however, show that $n > 2$ is optimal for identifying one uniquely bright peak in the pulse-profile, while $n \lesssim 2$ typically prefer a series of lower-amplitude peaks. For the curves corresponding to each value of $n$, we fit a complex polynomial and take its peak as the structure-maximizing DM. To estimate uncertainties, we fit a simple Gaussian function to the most prominent peak in the polynomial fit and take the 1$\sigma$ offset. For both $n=2$ and $n=4$, we find a structure-optimizing DM of $706.7^{+0.6}_{-0.6}~\mathrm{pc~cm}^{-3}$, as both contain an unambiguous peak.

\begin{figure*}
    \centering
    \hspace{-0.8cm}
    \includegraphics[width = 0.8 \textwidth]{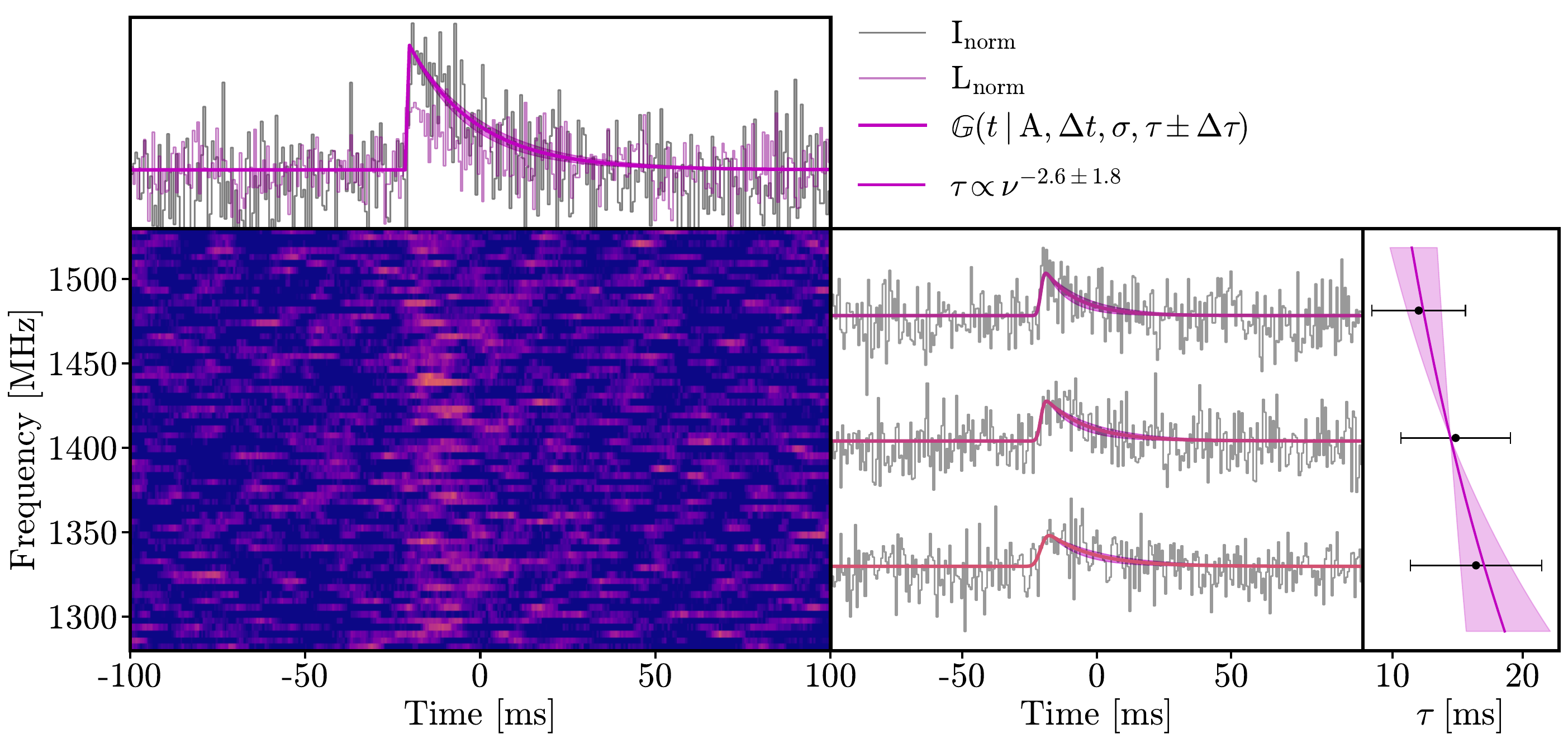}
    \caption{The dynamic (time-frequency) spectrum showing FRB\,20221219A, downsampled in time by a factor of 18 and in frequency by a factor of 96 to for clarity. To better visualize the burst in the dynamic spectrum, we further smoothed the data using a Savitzky-Golay smoothing filter with a window of 13 time bins and a polynomial order of 1. Above the dynamic spectrum, we show the normalized, unsmoothed total-intensity ($I_{\rm norm}$) and linearly polarized intensity ($L_{\rm norm}$) burst timeseries, as well as a non-linear least-squares fit for $\mathbb{G}(t)$ (Eq.~\ref{eq:expgauss}) to the pulse profile, which indicates a $\tau = 19.2_{-2.7}^{+2.7}$ ms. In the right-most panels, we fit timeseries to three $\simeq 83$ MHz sub-bands, and the respective scattering timescales measured for each sub-band with errors with a best-fit power-law of $\tau \propto \nu^{-2.6 \pm 1.8}$ power law. Uncertainties for all fits of $\mathbb{G}(t)$ are shown as shaded regions surrounding the fitted profiles.}
    \label{fig:nihariwfall}
\end{figure*}

Due to the low S/N of the burst, we confirm the presence of scattering by dividing the observing band into three $\simeq 83$ MHz sub-bands and fit exponentially-modified Gaussian functions $\mathbb{G}(t)$ to the timeseries of each respective sub-band, defined as

\begin{equation}\label{eq:expgauss}
\begin{aligned}
& \mathbb{G}\left(t \mid \mathrm{A}, \Delta t, \sigma, \tau\right)=\mathrm{A} \times\left[\exp \left(-\frac{\left(t-\Delta t\right)^2}{2 \sigma^2}\right)\right] \\
& \ast \left[\mathbb{H}\left(t-\Delta t\right) \exp \left(-\frac{t-\Delta t}{\tau}\right)\right],
\end{aligned}
\end{equation}

\noindent
where $t$ represents time, $\tau$ indicates the scattering timescale at the central frequency of the band (in MHz), $\mathbb{H}(t)$ is the Heaviside unit step function, $\Delta t$ is the time shift, $\sigma$ is Gaussian standard deviation, and $\ast$ symbolizes a convolution. We find that the pulse broadening timescale evolves in accordance with $\alpha = 2.6 \pm 1.8$, consistent with $\alpha \gtrsim 4$ scatter-broadening to within uncertainties.
Fitting $\mathbb{G}(t)$ to the timeseries obtained by integrating across the full observing band, we infer a scattering timescale $\tau = 19.2_{-2.7}^{+2.7}$ ms. The uncertainties in $\tau$ are derived from the covariance matrix of the least-squares fit, which is re-calculated for a set of pulse profiles within the DM uncertainties to produce a total error that includes uncertainty in DM. These fits are shown in Figure~\ref{fig:nihariwfall}.

\subsection{Optical/IR Follow-Up of Host \& Intervening Galaxies}\label{sec:hostID}

Using the 10$^{\mathrm{th}}$ data release of the Legacy Survey\footnote{https://www.legacysurvey.org/dr10/} \citep[DR10 R-band image;][]{Dey2019} and the Pan-STARRS1 \citep[PS1;][]{Chambers2016} optical survey, we identify a plausible host galaxy for FRB\,20221219A (henceforth HG\,20221219A) consistent with the radio localization ellipse (see R-band field in Figure~\ref{fig:niharidesifield}). We also identify two closely neighboring galaxies (henceforth IGH1 and IGH2).

We use \texttt{astropath} \citep{Aggarwal2021} for host galaxy validation, which implements a Bayesian approach to estimate association probabilities for neighboring galaxies with tunable priors based on the sky position of the FRB, as well as their R-band magnitude, position, and R-band radius (see SED fit in Figure~\ref{fig:niharised}; \citet{Sharma2023} and \citet{Law2023} provide details regarding host association methods). We were able to successfully associate the host galaxy to FRB\,20221219A at a confidence level of $99.6\%$ using Legacy Survey DR10 R-band imaging data from the Beijing-Arizona Sky Survey (BASS). We discuss IGH1 and IGH2, also visible in the field (see Figure~\ref{fig:niharidesifield}), in Sections \ref{sec:crowding}, \ref{sec:extmed}, and \ref{sec:cloud}.

\subsubsection{Optical/IR Photometry}

We make photometric measurements using images from PS1 ($g, r, i, z, y$), DECam \citep[$g, r, z$;][]{Valdes2014}, WISE \citep[$w1, w2$;][]{Wright2010}, as well as IR photometric data obtained with the Wide Field Infrared Camera \citep[WIRC, Ks;][]{Wilson2003} on August 16, 2022. HG\,20221219A, IGH1, and IGH2 are characterized following the methods outlined in \citet{Sharma2023}. We measure isophotes for each using PS1 i-band images, setting a best-fit coverage of $\gtrsim 95 \%$. The isophote is then scaled in accordance with the point spread functions of various photometric bands to execute aperture photometry.

\subsubsection{Optical/IR Spectroscopy}

We observed HG\,20221219A with the Deep Extragalactic Imaging Multi-Object Spectrograph \citep[Keck-II/DEIMOS;][]{Faber2003} on April 20, 2023, and both IGH1 and IGH2 with the Low-Resolution Imaging Spectrometer on the Keck I telescope \citep[Keck-I/LRIS;][]{Oke1995} on June 14, 2023. We created a mask of roughly 50 slits for our DEIMOS observation of the FRB\,20221219A field. Slits were placed on the positions of candidate foreground galaxies from the Legacy Survey DR9 catalog, including on massive central members of the galaxy cluster J171039.6+713427.

We reduce our spectra using \texttt{PypeIt} \citep{pypeit} and the standard \texttt{LPipe} software \citep{Perley2019} and calibrate using observations of the BD+28 4211 standard star. To account for slit losses, we scale the spectra to match the PS1 photometry.

To measure the spectroscopic redshifts and line fluxes, we use the penalized PiXel-Fitting software \citep[\texttt{pPXF;}][]{Cappellari2017, Cappellari2022}. This jointly fits for the stellar continuum and nebular emission lines based on the MILES stellar library \citep{SanchezBlazquez2006}. This places the host galaxy at a redshift of $z_{\mathrm{host}} = 0.554$, and find the two neighboring galaxies in the foreground at $z_{\mathrm{igh1}} = 0.492$ and $z_{\mathrm{igh2}} = 0.438$ (see Table~\ref{tab:hostintprops}).

\subsubsection{SED Modeling}

\begin{figure}
    \centering
    \hspace{-0.3 cm}
    \includegraphics[width = 0.47\textwidth]{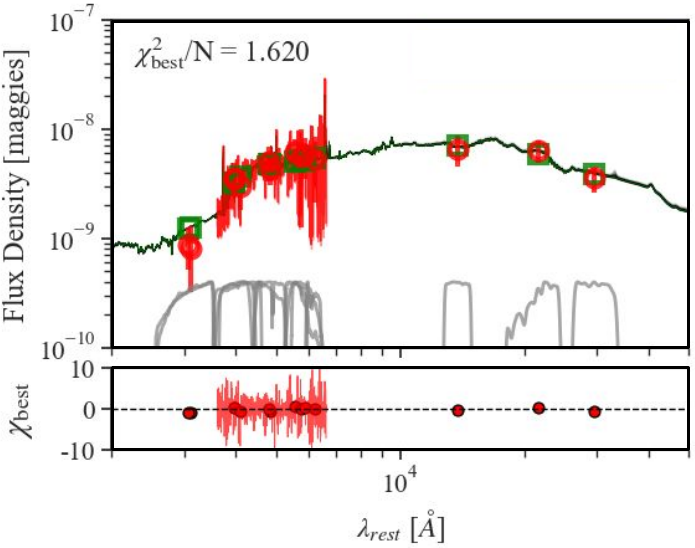}
    \caption{A non-parametric SED fit to observed spectroscopic and photometric data (red) for the host galaxy (PSO J$257.6335+71.625$) of FRB\,20221219A, obtained with Keck-II/DEIMOS. The SED fit is performed using \texttt{Prospector} with residuals for the best posterior sample (green).}
    \label{fig:niharised}
\end{figure}

Using the spectral energy distribution (SED) fitting software {\tt Prospector} \citep{Johnson2021}, we perform a non-parametric fit for the stellar properties of the HG\,20221219A, IGH1 and IGH2. While we were unable to obtain star formation rates for IGH1 and IGH2 due to low S/N in the FIR region of the spectrum, we were able to identify HG\,20221219A as a modestly star-forming galaxy, $\dot{M}_{\star} = 1.78_{-0.23}^{+0.24} ~M_\odot \mathrm{yr}^{-1}$, of stellar mass $\mathrm{log}_{10}(M_{\star, \mathrm{host}}) = 10.20^{+0.03}_{-0.04} ~M_\odot$ (see Figure~\ref{fig:niharised}), and IGH1, IGH2 as galaxies with stellar masses straddling that of HG\,20221219A, $\mathrm{log}_{10}(M_{\star, \mathrm{igh1}}) = 10.60^{+0.02}_{-0.02} ~M_\odot$, $\mathrm{log}_{10}(M_{\star, \mathrm{igh2}}) = 10.01^{+0.02}_{-0.02} ~M_\odot$ (see Table~\ref{tab:hostintprops}).

\subsection{Sightline Crowding}\label{sec:crowding}

\begin{figure}
    \centering
    \hspace{-1 cm}
    \hspace{0.88 cm}\includegraphics[width = 0.47\textwidth]{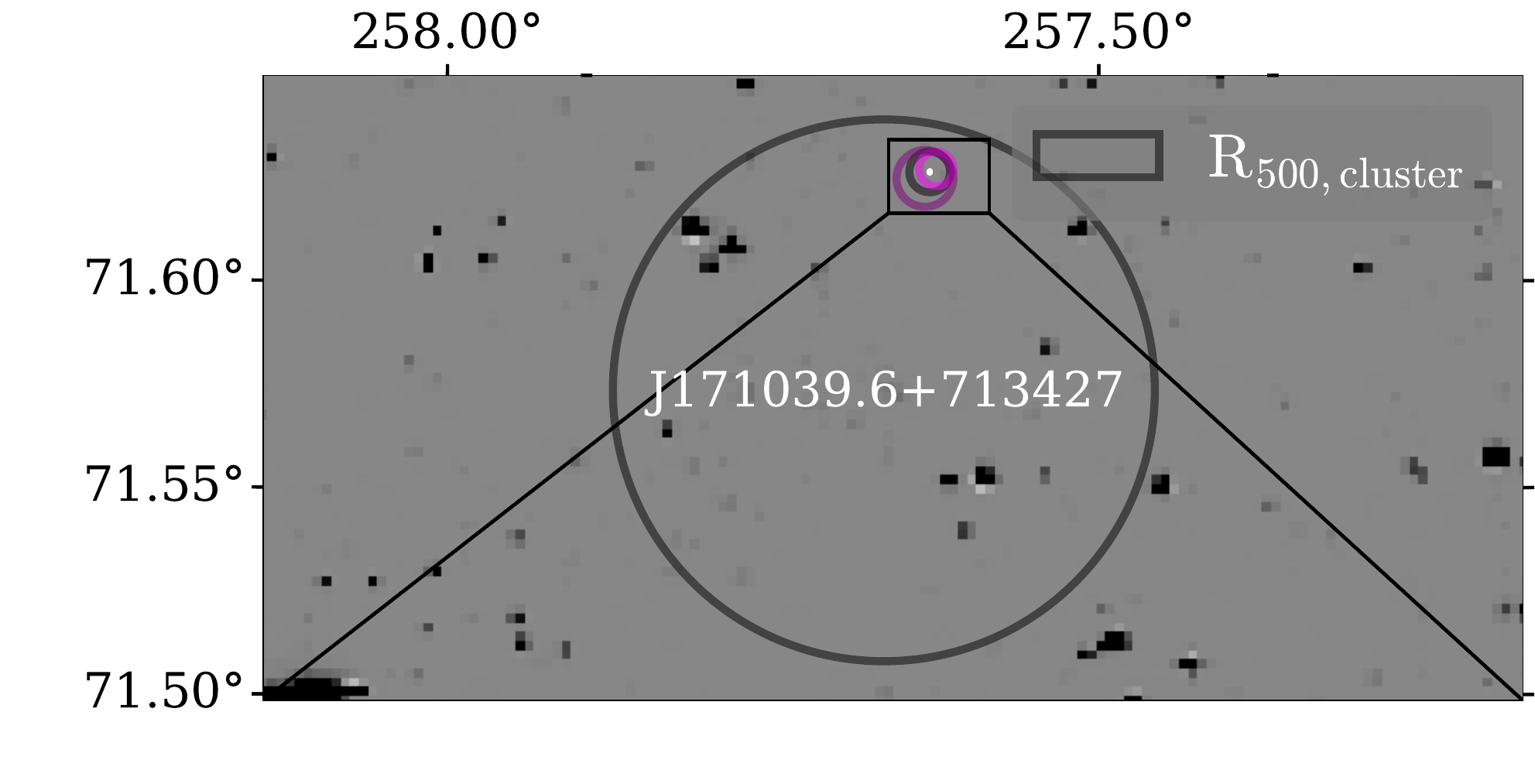}\vspace{-0.45 cm}
    \includegraphics[width = 0.47 \textwidth]{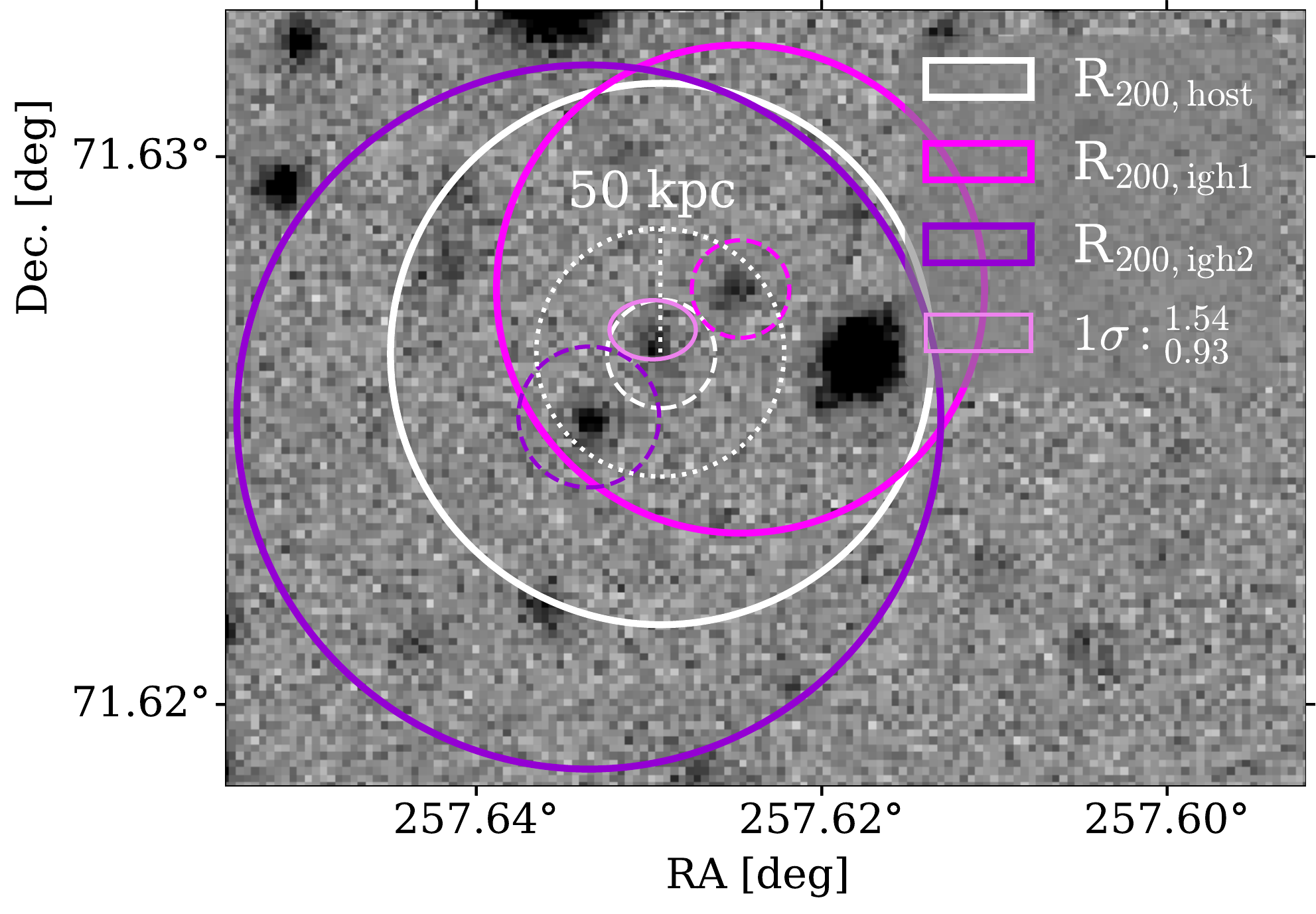}
    \caption{Legacy Survey (BASS) R-band cutout of the field surrounding FRB\,20221219A. The white dashed-line circle indicates a comoving radius of 50 kpc. The host and intervening galaxies are outlined by white, purple, and magenta dashed circles (labels consistent with Table~\ref{tab:hostintprops}). The virial radii ($\mathrm{R}_{200}$) are over-plotted as larger solid circles with consistent coloring. The 3$\sigma$ radio localization ellipse is over-plotted in violet. The upper panel shows the virial radius ($\mathrm{R}_{500}$) of the intervening galaxy cluster J171039.6+713427 identified in the WISE survey \citep{Wen2017}.}
    \label{fig:niharidesifield}
\end{figure}

The extreme degree of scattering in FRB\,20221219A and apparently unusual presence of closely neighboring intervening galaxies suggests that the galaxy population along the FRB\,20221219A LoS may be overcrowded. To evaluate whether this is true, we first calculate the number density of galaxies, $n_{50}(z)$, along the full sightline within a 50 kpc comoving radius using data taken from Legacy Survey DR10. A 50\,kpc radius is chosen as the smallest radius that fully encompasses the two interveners, accounting for uncertainty in the localization, as shown in Figure~\ref{fig:niharidesifield}; our conclusions are not sensitive to this choice. 

We then compare this count to the expected number densities using a Monte Carlo-based approach. The predicted number density of galaxies, $n_{50}(z)$, can be calculated using the halo mass function (HMF), which determines the number density $n$ for a specific halo mass $M_{h}$ as

\begin{equation}
    \frac{d n}{d \ln (M_{h})}=f\left(\sigma_m\right) \frac{\rho_{m, 0}}{M_{h}} \frac{d \ln \left(\sigma_m^{-1}\right)}{d \ln (M_{h})}
\end{equation}

\noindent
where $\rho_{m, 0}$ is the matter density at $z=0$, $\sigma_m$ is the root-mean-square variance of the linear density field, which varies with the linear matter power spectrum, and $f\left(\sigma_m\right)$ represents a function of $\sigma_m$ that is redshift-independent. We utilize the \citet{Tinker2008} HMF as implemented in the \texttt{hmf} Python package \citep{Murray2013}. The Tinker HMF is precisely calibrated to redshifts $z \lesssim 2$ and evolves with redshift in accordance with a predefined overdensity threshold.

To simulate expected values for $n_{50}(z)$ within the relevant spatial domain, we set a target redshift to that of the host, $z = 0.554$, and populate the comoving volume with dark matter halos following a Poisson distribution, based on the mean number of halos set by the HMF. Each halo is randomly assigned a mass in accordance with the HMF distribution, bounded by a mass range of $10^{11}~$-$~10^{15} M_\odot$. These limits are informed by the halo mass estimates for each intervener, $\log \left(M_{h, \mathrm{igh1}}\right) = 11.96^{+0.02}_{-0.02} M_\odot$, $\log \left(M_{h, \mathrm{igh2}}\right) = 11.52^{+0.01}_{-0.01} M_\odot$, which we obtain assuming the \citet{Moster2010} stellar-to-halo mass relation

\begin{equation}
\frac{M_\star}{M_h}(z) = 2 A(z)\left[\left(\frac{M_h}{M_A(z)}\right)^{-\beta(z)}+\left(\frac{M_h}{M_A(z)}\right)^{\gamma(z)}\right]^{-1},
\end{equation}

\noindent
parameterized by \citet{Girelli2020} using the $\Lambda$CDM DUSTGRAIN-\textit{pathfinder} simulation by $A = 0.0429_{-0.0026}^{+0.0026},\ M_A = 11.87_{-0.06}^{+0.06},\ \beta = 0.99_{-0.07}^{+0.08},\ \gamma = 0.669_{-0.015}^{+0.016}$ for a redshift range $0.2 \leq z \leq 0.5$ \citep[see Table 2 in][]{Girelli2020}.

For each simulation, we randomly select a sightline and calculate the number of halos within a 50 kpc comoving radius. Sightlines with a halo number density equal to or greater than the observed number density of two interveners are tallied, from which we construct a representative cumulative distribution function (CDF), thereby assessing the rarity of the sightline toward FRB\,20221219A. We find that for FRB\,20221219A CDF$\left(n_{50}(z) = 2\right) \sim 0.99$, indicating that the field is significantly overcrowded. The number of intersected halos that would be expected along this sightline amounts to only $N_{\mathrm{halos}} = 0.18$.

\begin{table*}
    \caption{Summary of observed and derived parameters for the host galaxy HG\,20221219A (see Figure~\ref{fig:niharised}), IGH1, and IGH2 (see Figure~\ref{fig:niharidesifield}). We report spectroscopic redshifts ($z$), impact parameters ($b$), stellar masses ($\log_{10} (M_\star)$), extinction-correct r-band magnitudes ($r$), and the star formation rate (SFR; $\dot{M}_{\star}$) of the host. The $\dot{M}_{\star}$ values of the interveners were not measured due to the low S/N of their respective SEDs.}
    \label{tab:hostintprops}
    \centering
    \begin{tabularx}{\textwidth}{@{\extracolsep{\fill}}XXXXXX}
        \hline \hline
        Object & $z$ & $b~\left[\mathrm{kpc}\right]$ & $\log_{10}\left(M_{\star}\right)~\left[M_\odot\right]$  & $r~\left[\mathrm{mag}\right]$ & $\dot{M}_{\star}~\left[M_{\odot}~\mathrm{yr}^{-1}\right]$ \\
        \hline
        HG\,20221219A & $0.554$ & \text{...} & $10.20_{-0.04}^{+0.03}$ & $22.6_{-0.2}^{+0.2}$ & $1.78_{-0.23}^{+0.24}$ \\
        IGH1 & $0.492$  & $43_{-11.3}^{+11.3}$ & $10.60_{-0.02}^{+0.02}$ & $21.6_{-0.2}^{+0.2}$ & \text{...} \\
        IGH2 & $0.438$ & $36.1_{-11.3}^{+11.3}$ & $10.01_{-0.02}^{+0.02}$ & $21.7_{-0.2}^{+0.2}$ & \text{...} \\
        \hline
    \end{tabularx}
\end{table*}

\begin{figure*}
    \centering
    \includegraphics[width = 0.99\textwidth]{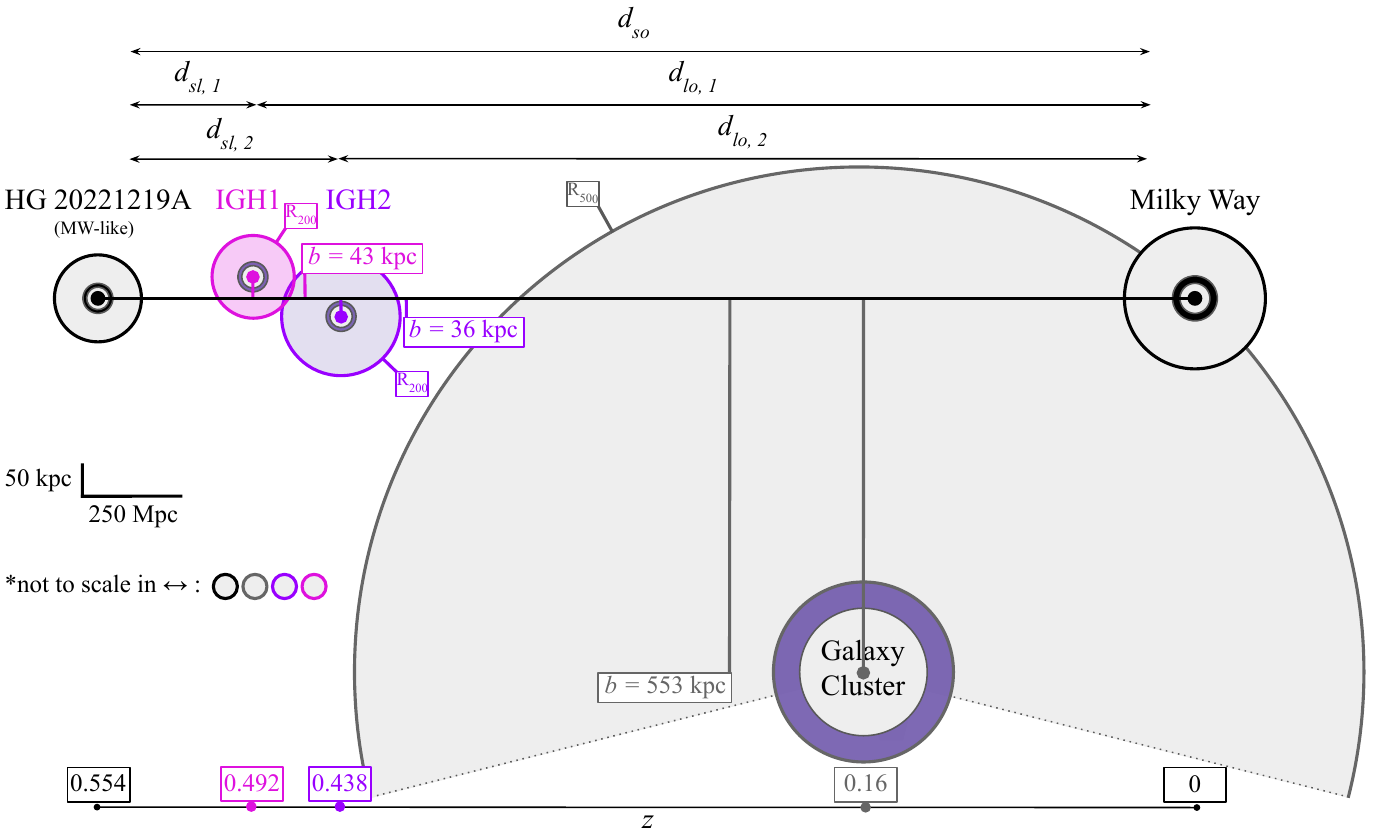}
    \caption{A schematic showing the FRB\,20221219A sightline, including the two intervening galaxies (IGH1, IGH2), their respective impact parameters ($b$), and virial radii ($\mathrm{R}_{200}$). We also show galaxy cluster J171039.6+713427 identified in the WISE survey \citep{Wen2017}, including its $\mathrm{R}_{500}$. The presence of this cluster is meaningful when considering the DM contribution of the ICM, discussed in Section~\ref{sec:dmbudget}.}
    \label{fig:niharilos}
\end{figure*}

\section{Analysis \& Discussion}

\subsection{The Dispersion Measure Budget} \label{sec:dmbudget}

The dispersion measure (DM) of the FRB is defined as the electron column density along the LoS $\mathrm{DM}=\int n_e(z) d l$, where $n_e$ is the electron density. The DM, expressed in standard units of $\mathrm{pc} \ \mathrm{cm}^{-3}$, is measured from the frequency-dependent delay in arrival times present in radio signals passing through cold, sparse plasmas along the LoS. As is typically done in the literature, we can express the measured DM for a source at redshift $z_{\mathrm{host}}$ as the sum

\begin{equation}\label{eq:dmbudget}
\begin{aligned}
\mathrm{DM}_{\mathrm{obs}}&=\mathrm{DM}_{\mathrm{mw}} + \mathrm{DM}_{\mathrm{igm}} \left(z_{\mathrm{host}}\right)\\
& +\frac{\mathrm{DM}_{\mathrm{icm}}}{1+z_{\mathrm{cluster}}} +\frac{\mathrm{DM}_{\mathrm{igh}}}{1+z_{\mathrm{igh}}}+\frac{\mathrm{DM}_{\mathrm{host}}}{1+z_{\mathrm{host}}}
\end{aligned}
\end{equation}

\noindent
that includes components from the Milky Way (mw), intergalactic medium (igm), intersecting intracluster medium (icm), intervening galaxy halos (igh), and host galaxy (host). Due to cosmic variance, a statistical dependence of $\mathrm{DM}_{\mathrm{igm}}$ on redshift, $z$, is needed. The $\mathrm{DM}_{\mathrm{icm}}$, $\mathrm{DM}_{\mathrm{igh}}$, and $\mathrm{DM}_{\mathrm{host}}$ terms require a further reduction of the rest-frame dispersion measures by factors of $1 /(1+z)$ for the respective redshifts of each. For simplicity, the host galaxy disk and halo, as well as the circumburst medium surrounding the FRB source, are treated as a single term. In the following sections, we will evaluate the DM contributions we expect from the Milky Way, IGM, ICM, and the intervening galaxies along the LoS. A schematic of the full sightline is shown in Figure~\ref{fig:niharilos}. 

Due to the presence of a closely neighboring galaxy cluster and two intervening galaxy halos, the DM budget is tight, leaving many of the diagnostics for DM outside the MW biased towards underestimates, as we describe later in \S\protect\ref{sec:igmdm} and \S\protect\ref{sec:ighdm}. To alleviate this bias and remain conservative in our DM attributions to local media, we omit the inclusion of a DM contribution from the MW halo, as this is also a largely uncertain quantity. The main purpose of this is to maximize the host galaxy DM allotment, which, if determined to be smaller, would aid in our suggestion that the observed scattering does not occur in the host. Upper limits on the DM of the Milky Way halo have, however, been constrained to within $\mathrm{DM}_{\mathrm{halo}} \lesssim 38$ pc cm$^{-3}$ by \citet{Ravi2023a} using the non-repeating FRB\,20220319D, and $\mathrm{DM}_{\mathrm{halo}} \lesssim 52~\text{-}~111$ pc cm$^{-3}$ by \citet{Cook2023} using the first FRB catalog from CHIME/FRB \citep{chime2021}.

\subsubsection{Milky Way and Halo}\label{sec:mwdm}

The DM contribution from the MW is characterized by the Galactic electron density model NE2001 \citep{Cordes2002}, which estimates $\mathrm{DM}_{\mathrm{mw}} \sim 44^{+4}_{-4} ~\mathrm{pc} \  \mathrm{cm}^{-3}$ \citep[assuming $\sim 10 \%$ uncertainty, in accordance with][]{Ocker2020} along the LoS toward FRB\,20221219A (positioned at $102.93^{\circ}$, $33.53^{\circ}$), through the Galaxy.

\subsubsection{Intergalactic Medium}\label{sec:igmdm}

While the IGM is a dominant contributor to the DM budgets of FRB signals \citep{Zhang2021, Walker2023}, it does not appear to noticeably scatter them. We assume this based on both modeling of gas densities in the IGM \citep{Macquart2013} and the absence of a universally applicable correlation between observed scattering timescales and extragalactic DMs \citep{chime2021, Chawla2022}. The mean DM contribution for a constant co-moving density in the IGM is given by \citep[e.g.,][]{McQuinn2014}

\begin{equation}\label{eq:dmigm}
\langle\mathrm{DM}_{\mathrm{igm}}(z)\rangle = n_{e_0} D_{\mathrm{H}} \int_0^z d z^{\prime} \frac{\left(1+z^{\prime}\right)}{E\left(z^{\prime}\right)}
\end{equation}

\noindent
where $n_{e_0}=2.2 \times 10^{-7} \times f_{\mathrm{IGM}}$ is the IGM electron density at $z=0$, $f_{\mathrm{IGM}}$ being the IGM baryon fraction based on Planck 18 cosmology \citep{Planck2020}, $D_{\mathrm{H}}=c / H_0$ is the Hubble distance and $E(z)=$ $\left[\Omega_{\mathrm{m}}(1+z)^3+1-\Omega_{\mathrm{m}}\right]^{1 / 2}$ for a flat $\Lambda$CDM universe with a matter density $\Omega_{\mathrm{m}}$. With this, we generate and fit a DM distribution using the probability density function (PDF) \citep{Macquart2020, Zhang2021}

\begin{equation}\label{eq:dmigm}
p_{\mathrm{igm}}(\Delta)=A \Delta^{-b} \exp \left[-\frac{\left(\Delta^{-a}-C_0\right)^2}{2 a^2 \sigma_{\mathrm{DM}}^2}\right], \quad \Delta>0
\end{equation}

\noindent
where $\Delta=\mathrm{DM}_{\mathrm{igm}} /\left\langle\mathrm{DM}_{\mathrm{igm}}\right\rangle$, $b$ depends on the halo gas density profile. We assume $a=b=3$ \citep{Macquart2020}. The effective standard deviation is $\sigma_{\mathrm{DM}}$, and $C_0$ impacts the transverse position, which is fitted for. We also assume values for $A$, $\sigma_{\mathrm{DM}}$, and $C_0$, derived from Illustris TNG simulation data by \citet{Zhang2021}. 

The IGM sightline for FRB\,20221219A is, however, slightly complicated by the presence of a proximate foreground galaxy cluster J171039.6+713427 cataloged in the WISE survey and DR9 group catalog, positioned at a redshift of $z_{\mathrm{cluster}}=0.16$ with mass $\log_{10}(M_{\mathrm{cluster}}) \simeq 14.1^{+0.2}_{-0.2} M_\odot$ (the quoted 0.2 dex uncertainty for cluster masses $\mathrm{log}_{10}\left(M_{\mathrm{cluster}}\right) \gtrsim 14 M_{\odot}$) and a LoS offset of $\simeq 3.2^{\prime}$, corresponding impact parameter $b \sim 553^{+11}_{-11}$ kpc \citep{Wen2017}. We evaluate the DM contribution of the intracluster medium (ICM) surrounding the cluster at its measured impacted parameter. The $n_{e, \mathrm{icm}}$ at $b$ is estimated using a Navarro-Frenk-White profile as implemented by \citet{Vikhlinin2006} and \citet{Prochaska2019}, contributing $\mathrm{DM}_{\mathrm{cluster}} \simeq 142^{+12}_{-12}$ pc cm$^{-3}$. This uncertainty does not, however, account for spatial asymmetry in the baryon halos, which may be significant \citep{Connor2023}.

\begin{figure}
    \centering
    \hspace{-0.3cm}
    \includegraphics[width = 0.48\textwidth]{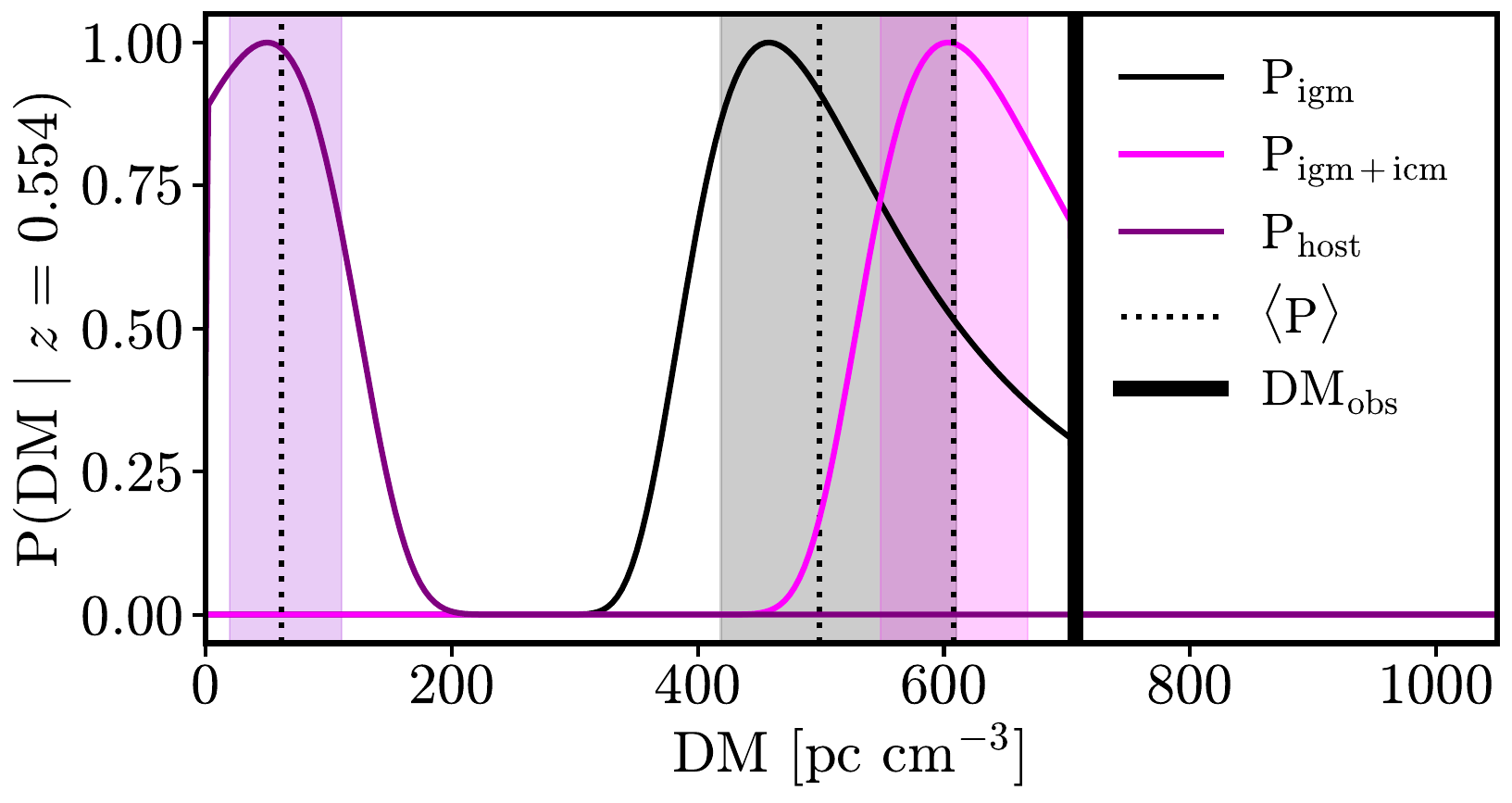}
    \caption{PDFs for DM$_{\mathrm{igm}}$ (black), DM$_{\mathrm{igm+cluster}}$ (magenta), DM$_{\mathrm{Host}}$ (purple), where P(DM $| ~z = 0.554$) is in units of [pc cm$^{-3}$]$^{-1}$. The 50$^{\mathrm{th}}$ percentile values of each PDF, as well as 15$^{\mathrm{th}}$ and 85$^{\mathrm{th}}$ percentile uncertainties are indicated by black dotted lines and shaded regions, respectively. Due to the narrowness of the Milky Way and intervening galaxy PDFs (see \S\protect\ref{sec:mwdm} and \S\protect\ref{sec:ighdm}), we omit them above for clarity.}
    \label{fig:dmigm}
\end{figure}

To estimate the collective PDF, P$_{\mathrm{igm + icm}}$(DM), for the DM contributions from the IGM and ICM, we assume a Gaussian PDF for $p_{\mathrm{icm}}(\Delta)$, given by

\begin{equation}\label{eq:gausspdf}
  p_{\mathrm{icm}}(\Delta)=\frac{1}{\sqrt{2 \pi \sigma_{\mathrm{DM}}^2}} \exp \left(-\frac{(\Delta-\mu)^2}{2 \sigma_{\mathrm{DM}}^2}\right)  
\end{equation}

\noindent
where $\Delta=\mathrm{DM}_{\mathrm{icm}} /\left\langle\mathrm{DM}_{\mathrm{icm}}\right\rangle$ and convolve with $p_{\mathrm{igm}}(\Delta)$ (see Eq.~\ref{eq:dmigm}) to obtain P$_{\mathrm{igm + icm}}$(DM). Taking the 50$^{\mathrm{th}}$ percentile of the convolution (uncertainties given by the 15$^{\mathrm{th}}$ and 85$^{\mathrm{th}}$ percentiles), we estimate $\mathrm{DM}_{\mathrm{igm+icm}, 50} = 608_{-59}^{+60} \mathrm{pc ~cm}^{-3}$ in the host frame, as shown in Figure~\ref{fig:dmigm} and Table~\ref{tab:dmpdfs}.



\subsubsection{Intervening Galaxies} \label{sec:ighdm}

We estimate DM$_{\mathrm{igh}}$ for the CGM of both intervening galaxies based on HII column densities (N$_{\mathrm{HII}}$) inferred from direct N$_{\mathrm{HI}}$ measurements made for similar nearby galaxies in the COS-Halos survey \cite[for 44 gaseous halos with $T \sim 10^{4}$ K photoionized CGMs at impact parameters of $b \lesssim 160$ kpc near $z \sim 0.2$;][]{Werk2014}. Inferences of N$_{\mathrm{HII}}$ rely on photoionization modeling relating N$_{\mathrm{HI}}$ to N$_{\mathrm{H}}$, which shows that the CGMs are highly ionized, such that $\mathrm{n}_{\mathrm{HII}} / \mathrm{n}_{\mathrm{H}} \gtrsim 99 \%$. Interpolating measurements from the COS-Halos survey across the full (N$_{\mathrm{HII}}$, $b$, $M_\star$) parameter space, we converge on the regions applicable to $b$ and $M_\star$ for both interveners and estimate $\mathrm{log}_{10}(\mathrm{N}_{\mathrm{HII}, \mathrm{igh1}}) \simeq 19.5^{+0.1}_{-0.1}$ cm$^{-2}$ and $\mathrm{log}_{10}(\mathrm{N}_{\mathrm{HII}, \mathrm{igh2}}) \simeq 19.6^{+0.1}_{-0.1}$ cm$^{-2}$, which correspond to $\langle\mathrm{DM}_{\mathrm{igh1}}(z =0.492)\rangle \simeq 10^{+5}_{-5} \mathrm{pc} \mathrm{~cm}^{-3}$ and $\langle\mathrm{DM}_{\mathrm{igh2}}(z =0.438)\rangle \simeq 14^{+5}_{-5} \mathrm{pc} \mathrm{~cm}^{-3}$. Similar to the ICM, we assume a Gaussian PDF for $p_{\mathrm{igh}}(\Delta)$. In estimating DM$_{\mathrm{Host}}$, we conservatively neglect the presence of any hot gas in the CGM of the interveners, which is again challenging to include in the DM budget due to tight constraints placed by the large contributions of the IGM and neighboring galaxy cluster. 

\subsubsection{Host Galaxy} \label{sec:hostdm}

To estimate DM$_{\mathrm{host}}$, we convolve $p_{\mathrm{igh + icm}}$ with $p_{\mathrm{MW}}$, $p_{\mathrm{igh1}}$ and $p_{\mathrm{igh2}}$ to obtain $\mathrm{P}_{\mathrm{host}}$(DM). We assume Gaussian PDFs (similar to Eq.~\ref{eq:gausspdf}) for the MW and both interveners. Taking the 50$^{\mathrm{th}}$ percentile of $\mathrm{P}_{\mathrm{host}}$(DM), we estimate $\mathrm{DM}_{\mathrm{host, 50}} = 62_{-43}^{+48}$ pc cm$^{-3}$ in the host-galaxy frame (see Figure~\ref{fig:dmigm} and Table~\ref{tab:dmpdfs}). Given the large uncertainties in modeling the DMs contributed by the intervening galaxy and cluster halos, we also estimated $\mathrm{DM}_{\mathrm{host, 50}}$ in the absence of their contributions. In that case, we find $\mathrm{DM}_{\mathrm{host, 50}} = 165_{-93}^{+81}$ pc cm$^{-3}$.


\begin{table}
    \centering
    \caption{Estimated DM contributions (in pc cm$^{-3}$) for each term in Eq.~\ref{eq:dmbudget} derived from the PDFs in Figure~\ref{fig:dmigm}, quoted as the 50$^{\mathrm{th}}$ percentile with 15$^{\mathrm{th}}$ and 85$^{\mathrm{th}}$ percentile uncertainties.}
    \label{tab:dmpdfs}
        \begin{tabularx}{0.47\textwidth}{@{\extracolsep{\fill}}XX} 
        \hline \hline
        $\mathrm{P(DM)}$ & $\mathrm{DM}_{50}$ $\left[\mathrm{pc~cm}^{-3}\right]$ \\
        \hline
        $\mathrm{MW}$ & $44^{+10}_{-10}$ \\
        $\mathrm{IGM + ICM}$ & $608_{-59}^{+60}$ \\
        $\mathrm{IGH1}$ & $10^{+5}_{-5}$ \\
        $\mathrm{IGH2}$ & $14^{+5}_{-5}$ \\
        $\mathrm{HG\,20221219A}$ & $62_{-43}^{+48}$ \\
        \hline
    \end{tabularx}
\end{table}

\subsection{The Scattering Budget} \label{sec:scatbudget}

We observe a high scattering timescale of $\tau_{\mathrm{obs}} = 19.2_{-2.7}^{+2.7}$ ms for FRB\,20221219A at 1.4 GHz. The pulse broadening, as measured for three individual sub-bands, is best described by $\tau \propto \nu^{-\alpha}$, $\alpha = 2.6 \pm 1.8$. To assess the origins of the scattering seen in FRB\,20221219A, we can divide $\tau_{\mathrm{obs}}$ for a source at redshift $z_{\mathrm{host}}$ into a sum of terms corresponding to the various scattering media along the LoS as

\begin{equation}\label{eq:taubudget}
\begin{aligned}
\tau(\nu)_{\mathrm{obs}}= & \tau_{\mathrm{mw}}(\nu)+\tau_{\mathrm{igm}}(\nu, z) ~+ \\
& \hspace{-1.5 cm}\frac{\tau_{\mathrm{igh}}(\nu)}{\left(1+z_{\mathrm{igh}}\right)^{3}}+\frac{\tau_{\mathrm{host}}(\nu)}{\left(1+z_{\mathrm{host}}\right)^{3}} ~ + \frac{\tau_{\mathrm{cbm}}(\nu)}{\left(1+z_{\mathrm{host}}\right)^{3}}
\end{aligned},
\end{equation}

\noindent that, similarly to the DM budget, includes lens rest-frame components from the Milky Way (mw), the intergalactic medium (igm), intervening galaxies (igh), the host galaxy (host), and now the circumburst medium (cbm). We adopt a nominal power-law frequency scaling, $\tau(\nu) \propto \nu^{-4}$. The last three terms of Eq.~\ref{eq:taubudget} scale with redshift to account for the fact that, for an observation frequency $\nu$, scattering goes as $\nu^{\prime}=\nu(1+z)$ in the galaxy's rest frame and time dilation goes as $(1+z)$ \citep{Macquart2013, Cordes2016}. In the following sections, we will consider each scattering medium along the line of sight individually.

\subsubsection{Electron Density Fluctuations}\label{sec:elecfluc}

Density fluctuations are modeled within a volume of ionized cloudlets, parameterized by the combined quantity $\widetilde{F}$, the density fluctuation parameter, defined as

\begin{equation}\label{eq:flucparam}
    \widetilde{F}=\frac{\zeta \varepsilon^2}{f_{\mathrm{v}}\left(l_{\mathrm{o}}^2 l_{\mathrm{i}}\right)^{1 / 3}} ~~~\left(\mathrm{pc}^2 \mathrm{~km}\right)^{-1 / 3},
\end{equation}

\noindent
where $l_{\mathrm{o}}$ is the outer scale of the turbulence power spectrum (in pc),  $l_{\mathrm{i}}$ is the inner scale (in km), $f_{\mathrm{v}}$ is the volume filling factor, $\epsilon^2$ is the intra-cloudlet variance in the electron density fluctuations, and $\zeta$ is the inter-cloudlet variation with respect to the mean electron density within the cloudlets. The fluctuation parameter $\widetilde{F}$, in combination with the geometric factor $G$, strongly affects $\tau$.

We assess the physical validity of two scattering scenarios: scattering by a uniform, extended CGM (\S\protect\ref{sec:extmed}), and scattering by a single ionized cloudlet (\S\protect\ref{sec:cloud}).

\subsubsection{Line of Sight Geometry}\label{sec:scatgeo}

In addition to the electron density fluctuations within intervening media, the position of those media with respect to the source and observer serves as an impactful factor in determining their scattering power. Treatments in \citet{Cordes2016}, \citet{Ocker2021}, and \citet{Cordes2022} define a dimensionless geometric factor $G$ to describe this. In Euclidean space, it is defined as

\begin{equation}
    G=\frac{\int_{\mathrm{layer}} s(1-s / d) ds }{\int_{\mathrm{host}} s(1-s / d) ds},
\end{equation}

\noindent
where $s$ represents a path element of the full path length $d$, and $G$ is of order unity for a source contained within in the scattering medium (e.g., the host galaxy), such that the distance to the source far exceeds the LoS extent of the scattering medium. Typically, however, $G$ strongly depends on the distribution of free electrons along the LoS and can increase by many orders of magnitude for intervening galaxies. In the case of FRB\,20221219A, the intervening galaxies we consider are at non-negligible redshifts, so the expression for $G$ can be rewritten (see \citet{Cordes2022} for the complete derivation) as

\begin{equation} \label{eq:cordesG}
    G\left(z_{\ell}, z_{\mathrm{s}}\right)=\frac{2 d_{\mathrm{sl}} d_{\mathrm{lo}}}{L d_{\mathrm{so}}},
\end{equation}

\noindent
where $d_{\mathrm{sl}}$, $d_{\mathrm{lo}}$, $d_{\mathrm{so}}$ are, respectively, the source-to-intervener, intervener-to-observer, and source-to-observer angular diameter distances (see schematic in Figure~\ref{fig:niharilos}), and $L$ is the thickness of the scattering layer. 

Given this geometric factor, a source at redshift $z_{\mathrm{s}}$, and a scattering region in an intervening galaxy at $z_{\ell}$, the scattering timescale can be expressed as

\begin{equation} \label{eq:cordesscat}
    \begin{aligned}
& \tau\left(\mathrm{DM}_{\ell}, \nu, z_{\ell}, z_{\mathrm{s}}\right) \\
& \quad \simeq 0.48 \mathrm{~ms} \times \frac{A_\tau \widetilde{F} G\left(z_{\ell}, z_{\mathrm{s}}\right) \mathrm{DM}_{\ell, 100}^2}{\nu^4\left(1+z_{\ell}\right)^3} ,
\end{aligned}
\end{equation}

\noindent
where $\mathrm{DM}_{\ell}$ is in the rest frame of the intervening galaxy, which contributes to the measured DM as $\mathrm{DM}_{\ell} /\left(1+z_{\ell}\right)$, with $\nu$ in $\mathrm{GHz}$ and $\mathrm{DM}_{\ell, 100} \equiv \mathrm{DM}_{\ell} /\left(100 ~\mathrm{pc} ~\mathrm{cm}^{-3}\right)$ \citep{Cordes2022}. The quantity $A_\tau$, in turn, accounts for the shape of the pulse broadening function (typically falling  between $1/6 ~\text{-}~ 1$, though we henceforth assume $A_\tau = 1$), and depends on both the spectral index $\beta$ and inner scale $l_{\mathrm{i}}$ of the scattering medium.

\subsubsection{Host Galaxy \& Circumburst Environment}\label{sec:hostscat}

It has been shown by \cite{Cordes2022} that the majority of FRBs detected by CHIME/FRB (repeating and non-repeating), assuming host galaxy-dominated scattering, emerge from an ISM with $\widetilde{F} \lesssim 1$ $\left(\mathrm{pc}^2 \mathrm{~km}\right)^{-1 / 3}$, with the exception of the heavily scattered FRB\,20191221A, which still suggests a host $\widetilde{F} \lesssim 10$ $\left(\mathrm{pc}^2 \mathrm{~km}\right)^{-1 / 3}$ \citep[see their Fig. 5;][]{Cordes2022}.

We find HG\,20221219A to be MW-like in its star formation rate (SFR; $\dot{M}_\star$). While the $M_{\star, \mathrm{host}}$ is a factor of $\sim 4$ smaller than that of the MW, and its SFR of $\dot{M}_{\star} = 1.78_{-0.23}^{+0.24} ~M_\odot \mathrm{yr}^{-1}$, is comparable to the MW's $\dot{M}_{\star} = 1.65^{+0.19}_{-0.19} ~M_\odot \mathrm{yr}^{-1}$ \citep{Licquia2015}. Under the assumption that the two galaxies are comparable in their properties, we can apply the established relation between $\tau$ and DM values of Galactic pulsars \citep{Cordes2016} to the host galaxy, and infer its scattering contribution based on DM$_{\mathrm{host}}$ estimated in Section~\ref{sec:hostdm}. First discussed by \citet{Sutton1971}, \citet{Rickett1977} and \citet{Cordes1991}, and most recently measured by \citet{Cordes2022}, this relation was fitted against scattering timescales and DMs for 568 Galactic pulsars. The canonical fitting function used for the pulsar data is $\widehat{\tau}(\mathrm{DM})=A \times \mathrm{DM}^a\left(1+B \times \mathrm{DM}^b\right)$ \citep{Ramachandran1997}, for which \citet{Cordes2016} found the empirical $\tau$-DM relation

\begin{equation}
\begin{aligned}
\left[\widehat{\tau}\left(\mathrm{DM}, \nu\right)\right]_{\mathrm{mw}, \mathrm{psr}} = 1.90 \times 10^{-7} \mathrm{~ms} \times \nu^{-\alpha} \mathrm{DM}^{1.5} \\
\times\left(1+3.55 \times 10^{-5} \mathrm{DM}^{3.0}\right),
\end{aligned}
\end{equation}

\noindent
at frequencies $\nu$ in $\mathrm{GHz}$, with scatter $\sigma_{\log \tau}=0.76$ dex \citep{Bhat2004, Cordes2019}. 

\begin{figure}
    \centering
    \hspace{-0.3cm}
    \includegraphics[width = 0.48\textwidth]{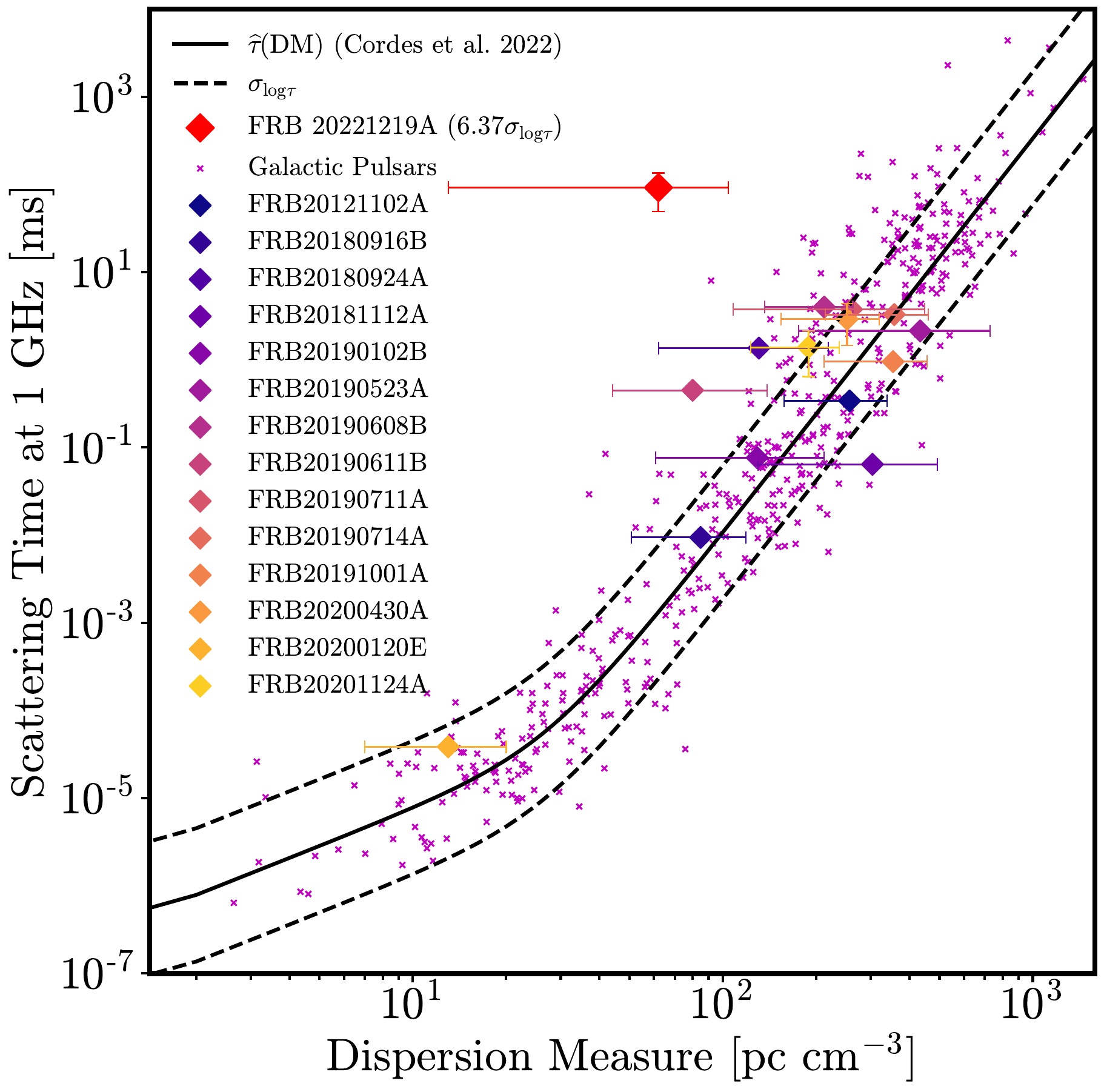}
    \caption{$\tau$-DM relation for Galactic pulsars. The fitted line (solid black) and $\pm 1 \sigma$ variations (dashed black) are based on measurements and upper limits on $\tau$ for Galactic pulsars \citep[568 in total; data provided by courtesy of J. Cordes and also published in][]{Cordes2016}. The scattering timescales and inferred host galaxy DMs for a subset of well-localized FRBs are over-plotted as multi-colored diamonds \citep{Cordes2022}. All scattering timescales have been scaled to their nominal values at 1 GHz. FRB\,20221219A (denoted by a red diamond) dramatically exceeds the expected scattering timescale from a Milky Way-like galaxy by 6.37$\sigma_{\mathrm{log}_{10}\tau}$ for its DM$_{\mathrm{host}}$ (see \S\protect\ref{sec:hostdm}).}
    \label{fig:tau-DM}
\end{figure}

Figure~\ref{fig:tau-DM} shows the fitted $\tau$-DM relation from \citet{Cordes2016} for Galactic pulsars. The steepening at large DMs occurs due to the cloudy nature of the inner Galaxy, where high-DM pulsars are found, as opposed to those in the outer regions of the galaxy or near the solar neighborhood \citep{Cordes1991, Cordes2019}. We over-plot a sample of well-localized FRBs in comparison
whose host DMs are inferred probabilistically using similar methods as those outlined in \S\protect\ref{sec:dmbudget}. FRB\,20221219A emerges from these populations as a clearly \textit{over-scattered} outlier\footnote{FRB scattering timescales have scaled up by a factor 3, under the assumption that the scattered waves are planar and not spherical \citep{Cordes2016}.} (to 6.37$\sigma_{\mathrm{log}_{10}\tau}$ for $[\widehat{\tau}(\mathrm{DM}, \nu)]_{\mathrm{MW}, \mathrm{psr}}$) given its inferred DM$_{\mathrm{host}}$, reflecting just how unique this event is in the context of most FRBs that have been detected to date. 
We, therefore, tentatively suggest that the host is unlikely to contribute significantly to the observed scattering timescale. Still, there remains the possibility that a compact scattering structure (e.g., a plasma lens) exists within HG\,20221219A and intersects the LoS. Such a structure could lead to strong deviations from the $\tau$-DM relation and weaken the inference made on the scattering prospects within HG\,20221219A. As this scenario is difficult to constrain without scintillation information, we do not expand on it in this work.

One possible indicator of a dynamic circumburst medium is a high Faraday rotation measure (RM), defined as $\mathrm{RM}=0.81 \int B_{\|}(l) n_{\mathrm{e}}(l) \mathrm{d} l$, where $B_{\|}$ is the magnetic field component along the LoS (in $\mu \mathrm{G}$), and $n_{\mathrm{e}}$ is the electron number density. The low S/N of FRB\,20221219A precludes any robust constraints on the RM by RM synthesis or $QU$-fitting techniques. We were, however, able to estimate a linear polarization fraction ($L/I$) in the non-derotated spectrum, which we take to be the lower limit. With this, we set an upper limit on its absolute $|\mathrm{RM}|$ via a Monte Carlo-based approach to estimate the degree to which a 100\% linearly polarized burst could be Faraday depolarized.

We calculate the likelihood of a ``true'' (intrinsic) $L/I$ exceeding the measured polarization fraction of FRB\,20221219A ($L/I \sim 0.502$) by simulating a linearly polarized signal completely in Stokes $Q$, a priori. Using off-pulse noise statistics, we proceed to simulate $Q$ and $U$ for a range of hypothetical RM values $|\mathrm{RM}|_{i}$ at the observing wavelength $\lambda$ as

\begin{equation}
\begin{aligned}
Q & =\cos \left(|\mathrm{RM}|_i \times \lambda^2\right) \\
U & =-\sin \left(|\mathrm{RM}|_i \times \lambda^2\right),
\end{aligned}
\end{equation}

\noindent
which we then sum in quadrature to obtain a linear polarization fraction $L/I = \sqrt{\langle Q \rangle^{2} + \langle U \rangle^{2}}$, where $\langle Q \rangle$ and $\langle U \rangle$ represent the mean of the simulated values over the signal data. From this distribution of simulated polarization fractions exceeding the measured value, we extract a lower limit $L/I \sim 0.42$ (estimated at $\sim$ 10\%) for FRB\,20221219A. Thus, we can set an upper limit on  $|\mathrm{RM}|$ of $\lesssim 345$ rad m$^{-2}$, a relatively low value, and conclude that the circumburst medium for FRB\,20221219A is not highly magnetic.

There is some evidence that a high RM is accompanied by strong local scattering \citep[such as in FRB\,20190520B;][]{Ocker2022b, AnnaThomas2023}. Since RM is a tracer of both $B_{\|}$ and $n_{\mathrm{e}}$, it is possible for a low RM to imply a less extreme local plasma environment and thus a lack of scattering power, as further evidenced by the high RMs of Galactic pulsars in the inner disk of the Milky Way, which exhibit high degrees of scattering \citep{Lazio1998, Wharton2012, Eatough2013, Cordes2022}. We note, however, that there is not necessarily correspondence between the $B_{\|}$ traced by RM and the possibly turbulent electron density fluctuations that lead to scattering. FRB\,20121102A, for instance, exhibits RM values $\gtrsim 10^{5}$ rad m$^{-2}$, indicating the presence of an extreme magneto-ionic local environment, but exhibits no significant scattering in its host or circumburst environment \citep{Michilli2018}. 

To further assess the likelihood that scattering would originate from the circumburst environment, we can take a more empirical approach using our combined parameters outlined in \S\protect\ref{sec:elecfluc} and \S\protect\ref{sec:scatgeo} to compute the measurable parameters, namely $G$, $\tau$, $\nu$, and $z_{\ell}$, and solve for $\widetilde{F} \times \mathrm{DM}_{\ell, 100}^{2}$ to better constrain the electron column density and fluctuation statistics. If we consider the scenario in which all the observed scattering originates from the circumburst environment (where $G = 1$), we find that $\widetilde{F} \times \mathrm{DM}_{\ell, 100}^2 \simeq 580 ~\mathrm{pc}^{4 / 3}\mathrm{~km}^{-1 / 3} \mathrm{~cm}^{-1 / 3}$. If we remain agnostic to the exact physics in the circumburst environment, and simply consider the lower limit on $\widetilde{F}$ imposed by DM$_{\mathrm{host}}$ (see Table~\ref{tab:dmpdfs}), we find that $\widetilde{F} \gtrsim 1500$ $\left(\mathrm{pc}^2 \mathrm{~km}\right)^{-1 / 3}$. While the exact range of $\widetilde{F}$ is poorly understood for local environments of FRB sources, we see that the lower limit far exceeds the maximal expected $\widetilde{F}$ for the thin disk ISM of a spiral galaxy by $\sim 10^{3}$ \citep[$\sim 10^{4}$ for dwarf galaxies, $\sim 10^{6}$ for elliptical galaxies;][]{Ocker2022c}. This assumption is purely illustrative, however, as it of course ignores the presence of an ISM in the host galaxy. Nonetheless, it highlights the requirement that the local environment be extremely turbulent in order for it to substantially contribute to $\tau_{\mathrm{obs}}$. 

Next, we'll evaluate the scattering contributions of each intervening galaxy, again utilizing the scattering formalism developed by \citet{Cordes2016}, \citet{Ocker2021}, and \citet{Cordes2022} to characterize the scattering power of intervening plasma based on the LoS geometry (\S\protect\ref{sec:scatgeo}), and electron density-fluctuation (i.e., turbulence) statistics (\S\protect\ref{sec:elecfluc}).





\subsubsection{IGH Model A: Extended CGM}\label{sec:extmed}

To evaluate the scattering power of a uniform, extended CGM, we assume DM$_{\mathrm{igh}}$ and a viral radius, $R_{200}$, based on the halo mass of each intervener. With these assumptions, we take a similar approach to \S\protect\ref{sec:hostscat} and solve for $\widetilde{F} \times \mathrm{DM}_{\ell, 100}^{2}$. First, we take the path length $L$ through the CGM (scattering layer) to be the geometric chord length through a spherically symmetric dark matter halo of $R_{200}$, corresponding to IGH1 or IGH2. With this, we calculate the geometrical factor $G$ using Eq.~\ref{eq:cordesG} for the angular diameter distances shown in Figure~\ref{fig:niharilos}. Substituting the DM values inferred for IGH1 and IGH2 in \S\protect\ref{sec:ighdm} based on the COS-Halos survey data \citep{Werk2014}, we estimate the value of $\widetilde{F}$ that would be required to reach $\tau_{\mathrm{obs}}$ using Eq.~\ref{eq:cordesscat} (see Table~\ref{tab:multimed}). We find the fluctuation parameters for each respective intervener, $\widetilde{F} \sim \left[1100^{+3300}_{-610}, 410^{+580}_{-190}\right]$ $\left(\mathrm{pc}^2 \mathrm{~km}\right)^{-1 / 3}$, that would be required to reach $\tau_{\mathrm{obs}}$ far exceed the physically valid range expected for the CGM based on prior studies \citep[$\tilde{F} \lesssim 10^{-3}$ $\left(\mathrm{pc}^2 \mathrm{~km}\right)^{-1 / 3}$;][]{Ocker2021}, requiring values more appropriate for the Galactic thin disk \cite[$\widetilde{F} \gtrsim 1$ $\left(\mathrm{pc}^2 \mathrm{~km}\right)^{-1 / 3}$;][]{Ocker2021}. 
 The specific range of predicted $\widetilde{F}$ is largely informed by the volume filling fraction of the gas, estimated to be $f_\mathrm{v} \sim 10^{-4}$ \citep{Vedantham2018}, inferred from areal covering fractions measured using quasar (QSO) absorption spectroscopy and photoionization modeling \citep{Prochaska2008, Stocke2013, Hennawi2015, Lau2016, McCourt2017}. The volume filling fraction relies on the assumption that the CGM can be described as a two-phase medium: a warm $T \gtrsim 10^{6}$ K medium of tenuous coronal gas that hosts within it a cooler $T \lesssim 10^{6}$ K medium, which becomes unstable around $T \sim 10^{4}$ K and fragments or shatters into discrete photoionized cloudlets that populate the tenuous gas with a high covering fraction, but low filling factor \citep{McCourt2017}. We conclude that scattering by an extended CGM is physically unreasonable.

This conclusion is, of course, highly sensitive to our estimate of $G$, and therefore $L_{\mathrm{halo}}$. Minimizing $L_{\mathrm{halo}}$ to by a factor of $\sim 10^{2}$ to of order $\sim 1 ~\mathrm{kpc}$, while keeping $\mathrm{DM}_{\mathrm{halo}}$ fixed, while an unphysical assumption, would better accommodate the scenario in which $\tau_{\mathrm{obs}}$ originates from an extended CGM. This illustrates that in order for scattering to occur in an IGH at all, it would need to originate from a highly confined scattering layer in the halo. We consider this possibility as follows in \S\protect\ref{sec:cloud}.

\begin{table}
    \centering
    \caption{For each IGH, we estimate halo masses ($\log_{10}\left(M_{h}\right)$), virial radii ($R_{200}$), impact parameters ($b$), geometrical chord lengths through the spherical halos ($L_{\mathrm{halo}}$). We then calculate geometrical factors ($G$; Eq.~\ref{eq:cordesG}) and solve for $\widetilde{F} \times \mathrm{DM}_{\ell, 100}^2$ (with units $\mathrm{pc}^{4 / 3}\mathrm{~km}^{-1 / 3} \mathrm{~cm}^{-1 / 3}$, using Eq.~\ref{eq:cordesscat}). For IGH model A, we infer $\mathrm{DM}_{\mathrm{igh}}$ using the COS-Halos survey data \citep[described in \S\protect\ref{sec:ighdm};][]{Werk2014}, and consequent values for $\widetilde{F}$. For IGH model B, we assume a cloudlet width ($L_{\mathrm{cloud}}$, which we take to be the outer scale $l_{o}$), electron density fluctuation statistics ($\zeta \varepsilon^2$), the Fresnel scale ($r_{F}$, which we take to be $10\times$ the inner scale $l_{i}$), and a volume-filling factor ($f_{v}$), to estimate the fluctuation parameter ($\widetilde{F}$), and infer a $\mathrm{DM}_{\mathrm{cloud}}$ required to achieve $\tau_{\mathrm{obs}}$. These assumptions are described and motivated in \S\protect\ref{sec:cloud}. All assumed or inferred parameters are indicated in bold font to distinguish them from data-driven measurements.}
    \label{tab:multimed}
    \begin{tabularx}{0.47\textwidth}{@{\extracolsep{\fill}}XXX} 
    \hline \hline
     Parameter & IGH1 & IGH2 \\
     \hline
      $\log_{10}\left(M_{h}\right) \left[M_\odot\right]$ & $11.52^{+0.05}_{-0.05}$ & $11.96^{+0.01}_{-0.01}$ \\
      $\mathrm{R}_{200}$ [kpc] & $122^{+5}_{-5}$ & $174^{+2}_{-2}$ \\
      $b$ [kpc] & $43^{+11}_{-11}$ & $36^{+11}_{-11}$ \\\\ 
      \hline
      \hline
      $\mathrm{~~~~~~~~~~~~~~~~\textit{IGH ~Model ~A: ~Extended ~CGM}}$ \\
      \hline
      $L_{\mathrm{halo}}$ [kpc] & $225^{+14}_{-14}$ & $340^{+5}_{-5}$ \\
      $G$ & $1100^{+70}_{-70}$ & $1300^{+19}_{-19}$ \\
      $\widetilde{F} \times \mathrm{DM}_{\ell, 100}^2$ & $11^{+2}_{-2}$ & 
      $8^{+1}_{-1}$ \\  
      $\mathrm{\textbf{DM}}_{\mathrm{\textbf{igh}}} \left[\mathrm{pc ~cm}^{-3}\right]$ & $10^{+5}_{-5}$ & $14^{+5}_{-5}$ \\
      $\widetilde{\bm{F}} \left[\left(\mathrm{pc}^2 \mathrm{~km}\right)^{-1 / 3}\right]$ & $1100^{+3300}_{-610}$ & $410^{+580}_{-190}$\\\\
      \hline
      \hline
      $\mathrm{~~~~~\textit{IGH ~Model ~B: ~Partially ~Ionized ~CGM ~Cloudlet}}$ \\
      \hline
      $\bm{L_{\mathrm{cloud}}}$ [$l_{o}$; pc] & 10 & 10 \\
      $\bm{G}$ & $2.4 \times 10^{7}$ & $4.4 \times 10^{7}$\\
      $\widetilde{\bm{F}} \times \bm{\mathrm{DM}_{\ell, 100}^2}$ & $5^{+0.7}_{-0.7} \times 10^{-4}$ & $2^{+0.3}_{-0.3} \times 10^{-4}$ \\
      $\bm{r_{\mathrm{F}}}$ [$10 \times l_{i}$; km] & $3.6 \times 10^{8}$ & $4.8 \times 10^{8}$ \\
      $\bm{\zeta \varepsilon^2}$ & 1 & 1 \\
      $\bm{f_{\mathrm{v}}}$ & $10^{-2}$ & $10^{-2}$ \\
      $\widetilde{\bm{F}} \left[\left(\mathrm{pc}^2 \mathrm{~km}\right)^{-1 / 3}\right]$ & $0.07$ & $0.06$\\
      $\bm{\mathrm{DM}_{\mathrm{cloud}}} \left[\mathrm{pc ~cm}^{-3}\right]$ & $8.5^{+0.5}_{-0.5}$ & $5.8^{+0.5}_{-0.5}$ \\  
      
      \hline
    \end{tabularx}
\end{table}

\subsubsection{IGH Model B: Partially Ionized CGM Cloudlet}\label{sec:cloud}

The likelihood of intersecting a partially ionized cloudlet can be understood based on the areal covering fraction of the cool ($T \sim 10^{4}$ K) gas. Multiple studies investigating the prevalence of absorption features in QSO spectra and lensing of those features \citep{Churchill2003, Prochaska2013, Lau2016}, as well as fluorescent Ly-$\alpha$ emission in halos \citep{Hennawi2015}, support the conclusion that the areal covering fraction exceeds unity. Hence, the scenario in which a single sightline intersects a cool cloudlet is highly probable.

To evaluate the scattering power of a single, partially ionized cloudlet, we again estimate $\widetilde{F} \times \mathrm{DM}_{\ell, 100}^2$  (Eq.~\ref{eq:flucparam}) based on a fiducial width of the cloudlet residing in an intervening CGM. The width is taken to be a typical upper limit on the size of a stable over-density of ionized gas in a cool ($T \sim 10^{4}$ K), fragmented CGM, $L_{\mathrm{cloud}} \sim 10$ pc \citep{McCourt2017}. We then estimate $\widetilde{F}$, based on fiducial values for the metric that describes local and ensemble electron density variations $\zeta \varepsilon^2$, the volume-filling factor $f_{\mathrm{v}}$, as well as the inner ($l_{i}$) and outer ($l_{o}$) scales of the turbulence power spectrum within the cloudlet. Under the well-justified assumption of homogeneous and isotropic turbulence, we take $\zeta \varepsilon^2 \sim 1$ \citep{Spangler1990, Armstrong1995, Rickett2009}. We also assume $f_{\mathrm{v}} \sim 10^{-2}$ \citep[typical for $T \sim 10^{4} ~\mathrm{K}$ gas;][]{Ocker2021}. It ought to be noted, however, that this factor is largely uncertain for individual clouds due to the inability to resolve and characterize this gas at sub-pc scales.

We set the inner scale $l_{i}$ to be a fraction\footnote{This is a somewhat arbitrary choice and can be adjusted. We set this inner scale as $0.1 \times r_{\mathrm{F}}$ as it concretely satisfies the condition that, in order for scattering to occur, the inner scale of the medium must lie below $r_{\mathrm{F}}$.} of the Fresnel scale $r_{\mathrm{F}}$, in this case $l_{i} = 0.1 \times r_{\mathrm{F}}$, at the distance of the intervener, defined as \citep{Cordes2017}

\begin{equation}
    r_{\mathrm{F}}=\sqrt{\frac{\lambda d_{\mathrm{sl}} d_{\mathrm{lo}}}{2 \pi d_{\mathrm{so}}}}
\end{equation}

\noindent
and an outer scale $l_{o}$ consistent with the width of the cloudlet (here $L_{\mathrm{cloud}} \sim 10 ~\mathrm{pc}$. With these assumptions, we estimate $\widetilde{F}$ using Eq.~\ref{eq:flucparam}, and the DM required to produce $\tau_{\mathrm{obs}}$ in Table~\ref{tab:multimed}. Under the assumptions outlined in Table~\ref{tab:multimed}, we find that for IGH2, a DM$_{\mathrm{cloud}}$ of only $5.8^{+0.5}_{-0.5} ~\mathrm{pc ~cm}^{-3}$ is required to achieve $\tau_{\mathrm{obs}}$. 

While there are a variety of circumgalactic media that could provide this electron column density, the most well-understood in the context of the Milky Way are high-velocity clouds (HVCs). The Wisconsin H$\alpha$ Mapper \citep[WHAM;][]{Tufte1998, Tufte1999} has observed HVCs at high galactic latitudes. As the extinction at high latitudes is presumed to be minimal, the H$\alpha$ luminosity can be related directly to emission measure ($\mathrm{EM} = \int n_{\mathrm{e}}^2 d l$), for which WHAM report an $\mathrm{EM}=0.18 \mathrm{~cm}^{-6} \mathrm{pc}$. Assuming an HII temperature of $T \sim 8000$ K and a small ionization fraction, $\mathrm{N}_{\mathrm{HII}}/\mathrm{N}_{\mathrm{HI}} \sim 0.08$ (as HVCs are known to be primarily neutral, ionized in part by external radiation), the inferred HII column densities along an individual sightline reach an upper limit of N$_{\mathrm{HII}} \sim 2 \times 10^{19}$ cm$^{-2}$ \citep{Tufte1999}.
This column density corresponds to a dispersion measure of DM$_{\mathrm{HVC}} \simeq 6.5 ~\mathrm{pc} ~\mathrm{cm}^{-3}$, which exceeds the DM$_{\mathrm{cloud}} \simeq 5.8^{+0.5}_{-0.5} ~\mathrm{pc} ~\mathrm{cm}^{-3}$ sufficient to achieve $\tau_{\mathrm{obs}}$.


While most Galactic HVCs lie at extra-planar distances $\lesssim 10$ kpc \citep{Wakker2008, Thom2006, Thom2008}, there are notable exceptions, such as those present in the Magellanic Stream (a stream of HVCs traversing $\gtrsim 100^{\circ}$ over the Galactic South Pole), which can extend out to $\sim 55~\text{-}~150$ kpc \citep{Besla2010}. Streams of this kind are not necessarily unique to the Milky Way, as tidal interactions between more massive galaxies and their neighboring dwarfs are not uncommon, particularly at higher $z$. Hence, while the presence of HVCs extending to extraplanar distances of order $b$ for IGH1 or IGH2 is challenging to constrain and largely uncertain, it is certainly possible. Galactic and Local Group HI surveys have also placed a subpopulation of so-called ``compact'' HVCs out to distances of $\sim 40~\text{-}~80$ kpc from the Galactic disk, and $\sim 50$ kpc around Andromeda \citep{Putman2003, Pisano2007, Westmeier2007}. Similar, even smaller ``ultra-compact'' HVCs have been observed within the Local Group, though their associations with individual galaxies remain uncertain \citep{Adams2013}. Still, they remain interesting candidate sources of FRB scattering.

Absorption line spectroscopy of QSOs has illuminated the intricate structures and dynamic processes characterizing the CGMs of galaxies beyond the Milky Way as well. High-velocity cloud-like structures, akin to those observed in the Milky Way's halo, have been detected in the CGMs of other galaxies, showcasing a wide range of ionization states and physical scales. QSOs intersecting the CGMs of intervening galaxies reveal the presence of ionized cloudlets through the detection of species ranging from low ions like Si II and Mg II to highly ionized ions such as O VI and Ne VIII \citep{Fox2014, Prochaska2017}. These findings suggest that the CGM is a complex, multiphase medium where different regions exhibit varying degrees of ionization \citep{Werk2013, Tumlinson2017}. The column density distribution and the spatial coherence of QSO absorption features suggest that these ionized clouds range from tens to hundreds of parsecs across, similar to HVCs in the Milky Way \citep{Stocke2013, Prochaska2017}. 

The distributions of these ionized structures within the galactic halos are substantial, extending from several tens to hundreds of kiloparsecs. This vast reach underscores the CGM's role as a reservoir of baryonic material that can influence galaxy evolution \citep{Werk2014, Burchett2016}. Ionization models employed to interpret the absorption spectra indicate that some regions within these clouds are nearly fully ionized, revealing a complex CGM phase structure \citep{Oppenheimer2013, Stocke2013}. They also indicate that these clouds occupy diverse thermal phases of the CGM, from cool ($T \sim 10^{4}$ K), photoionized gas to hot ($T\gtrsim 10^{6}$ K), collisionally ionized plasma \citep{Werk2016, Tumlinson2017}.

The evidence garnered from spectroscopic studies not only confirms the presence of multiphase, dynamic HVC-like structures but also reveals their substantial physical scales and diverse ionization fractions, making it all the more reasonable that an intersection with a single, marginally ionized cloudlet could occur for both IGH1 and IGH2, given their respective impact parameters, and therefore dominate the scattering budget of FRB\,20221219A. 


\section{Conclusions} \label{sec:futurework}

By constructing detailed DM and scattering budgets for FRB\,20221219A, we find that the event is highly over-scattered in its host galaxy and lies along an unusually rich sightline. We also find that the IGM and ICM likely dominate the DM budget, leaving a modest allowance for DM$_{\mathrm{host}}$.

We consider the possibility of scattering occurring in the CGMs of two closely intervening galaxies, and find it plausible. Specifically, we note that a fortuitous intersection with a single, partially ionized cloudlet in one of the CGMs could produce $\tau_{\mathrm{obs}}$. The geometric leverage offered by compact scattering structures at large distances from both the source and observer, as well as the potential presence of partially ionized structures at the impact parameters of both IGH1 and IGH2 along the LoS, make this scenario likely. We further find that nothing about the host galaxy or local environment is unusual with respect to other FRBs (and pulsars) with far lower scattering, lending credence to this suggestion.

The fact that all FRBs are not scattered by the CGMs of intervening galaxies can be explained by the far sparser distributions of such galaxies along most sightlines, as well as the known bias against observing highly scattered FRBs. We further note that scattering is highly stochastic for fixed DMs, as evidenced by MW pulsars, inflating this bias even more.


In future case studies, it would be of great use, where possible, to measure scintillation bandwidths in addition to pulse broadening to better understand FRB scattering mechanisms and disentangle distinct LoS scattering media. Comparisons between scintillation and pulse broadening of a single burst can assist in constraining whether the scattering effects in FRBs are due to one or multiple screens along the line of sight. Still, it is crucial to search for repeat bursts from heavily scattered FRBs to characterize additional time-variable effects.


\section{Acknowledgements}

The authors would like to thank Jim Cordes for insightful conversations, as well as the staff members of the Owens Valley Radio Observatory and the Caltech radio group whose efforts were vital to the success of the DSA-110, including Kristen Bernasconi, Stephanie Cha-Ramos, Sarah Harnach, Tom Klinefelter, Lori McGraw, Corey Posner, Andres Rizo, Michael Virgin, Scott White, and Thomas Zentmyer. The DSA-110 is supported by the National Science Foundation Mid-Scale Innovations Program in Astronomical Sciences (MSIP) under grant AST-1836018. 

This research is based in part on data gathered with the W.M. Keck Observatory, which is operated, in scientific partnership, by the California Institute of Technology, the University of California, and the National Aeronautics and Space Administration. The Observatory was made possible by the generous financial support of the W.M. Keck foundation.

The Legacy Surveys consist of three individual and complementary projects: the Dark Energy Camera Legacy Survey (DECaLS; Proposal ID \#2014B-0404; PIs: David Schlegel and Arjun Dey), the Beijing-Arizona Sky Survey (BASS; NOAO Prop. ID \#2015A-0801; PIs: Zhou Xu and Xiaohui Fan), and the Mayall z-band Legacy Survey (MzLS; Prop. ID \#2016A-0453; PI: Arjun Dey). DECaLS, BASS and MzLS together include data obtained, respectively, at the Blanco telescope, Cerro Tololo Inter-American Observatory, NSF’s NOIRLab; the Bok telescope, Steward Observatory, University of Arizona; and the Mayall telescope, Kitt Peak National Observatory, NOIRLab. Pipeline processing and analyses of the data were supported by NOIRLab and the Lawrence Berkeley National Laboratory (LBNL). The Legacy Surveys project is honored to be permitted to conduct astronomical research on Iolkam Du’ag (Kitt Peak), a mountain with particular significance to the Tohono O’odham Nation.

NOIRLab is operated by the Association of Universities for Research in Astronomy (AURA) under a cooperative agreement with the National Science Foundation. LBNL is managed by the Regents of the University of California under contract to the U.S. Department of Energy.

This project used data obtained with the Dark Energy Camera (DECam), which was constructed by the Dark Energy Survey (DES) collaboration. Funding for the DES Projects has been provided by the U.S. Department of Energy, the U.S. National Science Foundation, the Ministry of Science and Education of Spain, the Science and Technology Facilities Council of the United Kingdom, the Higher Education Funding Council for England, the National Center for Supercomputing Applications at the University of Illinois at Urbana-Champaign, the Kavli Institute of Cosmological Physics at the University of Chicago, Center for Cosmology and Astro-Particle Physics at the Ohio State University, the Mitchell Institute for Fundamental Physics and Astronomy at Texas A\&M University, Financiadora de Estudos e Projetos, Fundacao Carlos Chagas Filho de Amparo, Financiadora de Estudos e Projetos, Fundacao Carlos Chagas Filho de Amparo a Pesquisa do Estado do Rio de Janeiro, Conselho Nacional de Desenvolvimento Cientifico e Tecnologico and the Ministerio da Ciencia, Tecnologia e Inovacao, the Deutsche Forschungsgemeinschaft and the Collaborating Institutions in the Dark Energy Survey. The Collaborating Institutions are Argonne National Laboratory, the University of California at Santa Cruz, the University of Cambridge, Centro de Investigaciones Energeticas, Medioambientales y Tecnologicas-Madrid, the University of Chicago, University College London, the DES-Brazil Consortium, the University of Edinburgh, the Eidgenossische Technische Hochschule (ETH) Zurich, Fermi National Accelerator Laboratory, the University of Illinois at Urbana-Champaign, the Institut de Ciencies de l’Espai (IEEC/CSIC), the Institut de Fisica d’Altes Energies, Lawrence Berkeley National Laboratory, the Ludwig Maximilians Universitat Munchen and the associated Excellence Cluster Universe, the University of Michigan, NSF’s NOIRLab, the University of Nottingham, the Ohio State University, the University of Pennsylvania, the University of Portsmouth, SLAC National Accelerator Laboratory, Stanford University, the University of Sussex, and Texas A\&M University.

BASS is a key project of the Telescope Access Program (TAP), which has been funded by the National Astronomical Observatories of China, the Chinese Academy of Sciences (the Strategic Priority Research Program “The Emergence of Cosmological Structures” Grant \#XDB09000000), and the Special Fund for Astronomy from the Ministry of Finance. The BASS is also supported by the External Cooperation Program of Chinese Academy of Sciences (Grant \#114A11KYSB20160057), and Chinese National Natural Science Foundation (Grant \#12120101003, \#11433005).

The Legacy Survey team makes use of data products from the Near-Earth Object Wide-field Infrared Survey Explorer (NEOWISE), which is a project of the Jet Propulsion Laboratory/California Institute of Technology. NEOWISE is funded by the National Aeronautics and Space Administration.

The Legacy Surveys imaging of the DESI footprint is supported by the Director, Office of Science, Office of High Energy Physics of the U.S. Department of Energy under Contract No. DE-AC02-05CH1123, by the National Energy Research Scientific Computing Center, a DOE Office of Science User Facility under the same contract; and by the U.S. National Science Foundation, Division of Astronomical Sciences under Contract No. AST-0950945 to NOAO.

The Pan-STARRS1 Surveys (PS1) and the PS1 public science archive have been made possible through contributions by the Institute for Astronomy, the University of Hawaii, the Pan-STARRS Project Office, the Max-Planck Society and its participating institutes, the Max Planck Institute for Astronomy, Heidelberg and the Max Planck Institute for Extraterrestrial Physics, Garching, The Johns Hopkins University, Durham University, the University of Edinburgh, the Queen's University Belfast, the Harvard-Smithsonian Center for Astrophysics, the Las Cumbres Observatory Global Telescope Network Incorporated, the National Central University of Taiwan, the Space Telescope Science Institute, the National Aeronautics and Space Administration under Grant No. NNX08AR22G issued through the Planetary Science Division of the NASA Science Mission Directorate, the National Science Foundation Grant No. AST-1238877, the University of Maryland, Eotvos Lorand University (ELTE), the Los Alamos National Laboratory, and the Gordon and Betty Moore Foundation.

\vspace{5mm}
\facilities{DSA-110, Keck-I/LRIS \citep{Oke1995}, Keck-II/DEIMOS \citep{Faber2003}}

\software{\texttt{astropy} \citep{astropy2018}; \texttt{scipy} \citep{scipy2020}, \texttt{numpy} \citep{numpy2020}, \texttt{hmf} \citep{hmf}, \texttt{matplotlib} \citep{matplotlib2007}, \texttt{LPipe} \citep{Perley2019}, \texttt{pPXF} \citep{Cappellari2017, Cappellari2022}, \texttt{PypeIt} \citep{pypeit}, \texttt{prospector} \citep{Johnson2021}, \texttt{astropath} \citep{Aggarwal2021}}

\bibliography{refs}
\bibliographystyle{aasjournal}

\end{document}